\documentstyle[11pt,aaspp4]{article}
\def\gtsim{\hbox{\raise.5ex\hbox{$>$}\llap{\lower.5ex\hbox{$\sim$}}}}
\def\ltsim{\hbox{\raise.5ex\hbox{$<$}\llap{\lower.5ex\hbox{$\sim$}}}}

\begin{document}

\title{Photometric Properties of 47 Clusters of Galaxies: \\ 
   I. The Butcher-Oemler Effect}

\author{V. E. Margoniner \altaffilmark {1,2} and R. R. de Carvalho 
\altaffilmark {2} }

\affil{\it{Email: vem@physics.bell-labs.com, reinaldo@on.br }}

\altaffiltext{1}{Bell Laboratories, Lucent Technologies, Murray Hill, NJ 07974}
\altaffiltext{2}{Observat\'orio Nacional, CEP 20921-400, Rio de Janeiro, Brazil}

\begin{abstract}

We present $gri$ CCD photometry of 44 Abell clusters and 4 cluster candidates.
Twenty one clusters in our sample have spectroscopic redshifts.
Fitting a relation between mean $g$, $r$ and $i$ magnitudes, and redshift 
for this subsample, we have calculated photometric redshifts for the remainder
with an estimated accuracy of $\sim 0.03$.
The resulting redshift range for the sample is $ 0.03 < z < 0.38$.
Color-magnitude diagrams are presented for the complete sample and used to
study evolution of the galaxy population in the cluster environment.
Our observations show a strong Butcher-Oemler effect (Butcher \& Oemler
1978, 1984), with an increase in the fraction of blue galaxies ($f_B$) with
redshift that seems more consistent with the steeper relation estimated by Rakos 
and Schombert (1995) than with the original one by Butcher \& Oemler (1984).  
However, in the redshift range between $\sim 0.08$ and $0.2$, where 
most of our clusters lie, there is a wide range of $f_B$ values, 
consistent with no redshift evolution of the cluster galaxy population. 
A large range of $f_B$ values is also seen between $\sim 0.2$ and $0.3$, 
when Smail at al. (1998) x-ray clusters are added to our sample.
The discrepancies between samples underscore the need for an unbiased sample
to understand how much of the Butcher-Oemler effect is due 
to evolution, and how much to selection effects. 
We also tested the idea proposed by Garilli et al. (1996) that there
is a population of unusually red galaxies which could be associated
either with the field or clusters, but we find that these
objects are all near the limiting magnitude of the images ($20.5 < r < 22$)
and have colors that are consistent with those expected for stars or field
galaxies at $z\sim 0.7$.

\end{abstract}

\keywords{galaxies: clusters -- evolution -- color -- photometry}

\clearpage

\section{INTRODUCTION}

The existence of a linear locus in the color-magnitude diagram is a
main characteristic of the early-type galaxy population in clusters. 
The effect in which the brightest galaxies are also the most red ones 
was studied in a systematic way by various authors. Visvanathan \&
Sandage (1977) obtained spectral energy distributions (SEDs) over a wide range
of wavelengths for Virgo cluster galaxies and were one of the first to 
establish not only the existence, but also the dependence of the slope 
of the color-magnitude relation with wavelength.

The slope of the color-magnitude relation can be explained as an age 
effect, implying that the most luminous, massive galaxies, are older, or 
as a metallicity effect, consistent with the most luminous galaxies being 
more metal rich. {\it Hubble Space Telescope} (HST) observations 
have shown that the color-magnitude 
relation continues to be very well defined for early-type galaxies in high 
redshift clusters up to $z\sim1$ (Standford et al. 1998, Kodama et al. 1998, 
Ellis et al. 1997), as well as in the field (Kodama et al. 1999). These 
observations suggest that at high redshift the early-type population already 
consists of old, passively evolving systems. Ellis et al. (1997), 
Standford et al. (1998) estimated that these galaxies were formed at  $z>2$, 
and therefore the color-magnitude relation can not be explained as a purely 
age effect.
Also, Terlevich et al. (1999) used spectral absorption line indices, together 
with broad band photometry, to investigate how mean age and metal abundance 
correlate with galaxy luminosity and find that the color-magnitude relation 
in Coma is driven primarily by a luminosity-metallicity correlation.

However, recent 
observations by Worthey (1996) indicate that star formation is 
still occurring in $\sim 2/3$ of low redshift ellipticals and that 
many of these galaxies have ages less than half a Hubble time. The scatter 
in age of the early-type galaxy population can still be consistent with a 
well defined color-magnitude relation if galaxies assembling more recently 
are on average more metal-rich than older galaxies of same luminosity 
(Ferreras et al. 1998), which is consistent with Worthey's observations 
that show a trend for younger large galaxies to be more metal rich. 

The luminosity-metallicity relation can be created as a result of the 
different efficiency of supernova galactic winds to eject gas from galaxies 
with different masses (Larson 1974). Massive galaxies are able to retain a high 
fraction of their gas, becoming more enriched than less massive galaxies which 
tend to lose their gas more easily. Another possible explanation for the origin 
of the luminosity-metallicity relation is the difference in merging 
dynamics between systems of different luminosities (Bekki and Shioya 1997). 
If the more luminous elliptical galaxies are formed by galaxy merging with 
more rapid star formation, less gas is tidally stripped from these systems 
during merging, and as a result a greater amount of the gas is enriched.

The universality of the color-magnitude relation (Visvanathan \& Sandage 1977, 
Bower et al. 1992) makes it a powerful tool to study characteristics of the 
cluster galaxy population and its evolution. Butcher and Oemler (1978) 
observed the central regions of two rich, high redshift clusters of galaxies 
(CL 3C 295 at z=0.46 and CL 0024+1654 at z=0.39) and found an excess of 
blue galaxies in comparison to the typical early-type population which 
is found in the central region of local clusters (Dressler 1980).
In 1984, Butcher \& Oemler published a study of 33 clusters in the 
redshift range between z=0.003 and 0.54 and confirmed that the fraction 
of blue galaxies in the inner region of clusters, where $30\%$ of their 
galaxy population is found, increases with redshift.
Butcher and Oemler's results were interpreted as detection of evolution in the 
population of galaxies, and many works regarding the nature of the blue 
galaxies followed.

Larson et al. (1980) suggested an evolutionary connection between S0 and 
spiral galaxies as the physical origin of the enhanced star formation in 
moderate redshift galaxies. The authors argue that S0s might be disk systems 
that lost their gas envelopes during the cluster collapse and consumed their 
remaining gas by star formation. This idea can explain the population of 
blue galaxies observed by Butcher and Oemler as spiral galaxies seen just 
before running out of gas, and the disappearance of this population in more 
evolved, low redshift rich clusters (Dressler et al. 1997, Couch et al. 1998). 
This idea is also in agreement with observations showing that the blue 
population lies preferentially in the outer edges of the cluster (Butcher 
and Oemler 1984, Rakos et al. 1997).
Also, early-type galaxies in dense clusters are less metal-enriched 
than their counterparts in lower density environments (Rose et al. 1994), 
suggesting the truncation of star-formation in high density regions. The idea 
that the Butcher-Oemler effect is an evolutionary phenomenon is also reinforced 
by observations showing that star formation similar to that seen in distant 
clusters is still ongoing, although at a reduced level, in local clusters 
(Caldwell and Rose 1997). 

HST images allowed the determination of the 
morphology of galaxies in high redshift clusters (Dressler et al. 1994,
Couch et al. 1994, 1998). These data suggest that the blue Butcher-Oemler 
galaxies are predominantly normal late-type (i.e. small B/D ratios) spirals 
but also that dynamical interactions and mergers between galaxies may be an 
important process responsible for the star formation enhancement in those 
galaxies. Oemler et al. (1997) presented HST obsevations of four rich 
clusters at $z\sim0.4$ confirming that most of the blue Butcher-Oemler 
galaxies have colors, luminosities and spatial distributions similar to 
the normal galaxies observed at low redshift. The authors show however that 
$\sim 30\%$ of these ``normal'' galaxies show abnormal patterns of star 
formation such as rings. Oemler et al. (1997) also find a large fraction 
of m/i galaxies, but argue that these objects can not account for most of 
the blue Buther-Oemler galaxies. m/i galaxies do not follow the spatial 
distribution of the the blue Buther-Oemler objects, instead they are 
concentrated at the dense regions of the cluster following the distribution 
observed for E/S0 galaxies. Furthermore, Oemler et al. (1997) show that the
m/i galaxies are too few to account for the Buther-Oemler galaxies. 
Rakos et al. (1996) used narrowband photometry to study Butcher-Oemler 
galaxies and also found that the blue population consists of spiral-like 
systems with the addition of a small fraction of bright, blue interacting 
and merger systems. Rakos and Schombert (1995) find that the fraction of 
blue galaxies increases from $20\%$ at $z=0.4$ to $80\%$ at $z=0.9$, 
suggesting that the evolution in clusters is even stronger than previously 
indicated by Butcher and Oemler.  

Although the Butcher-Oemler effect has been debated in many studies, 
and many evolutionary theories have been proposed, there are some
suggestions of strong selection biases in Butcher and Oemler's original sample of 
clusters. Newberry et al. (1988) measured velocity dispersions and 
surface density of galaxies in clusters  
and found that there is a marked difference between local clusters and 
intermediate redshift ones. More recently, Andreon and Ettori (1999) 
measured x-ray surface brightness profiles, size, and luminosity 
of the Butcher-Oemler sample of clusters and conclude that 
this sample is not formed from the same kind of objects observed over a 
range of look-back times. The selection effects are not well 
understood, and might be mimicking evolutionary effects. Smail et al. 
(1998) used a well defined sample of 10 clusters in the redshift range 
$z=0.22-0.28$ with high x-ray luminosity, and found that these massive 
clusters contain only a small fraction of blue galaxies. 
The Butcher-Oemler effect is not observed in this sample.

 Also, in a visually selected sample of clusters, the presence of false 
structures, identified as clusters due to projection effects, is expected. 
The color of the galaxy population at such regions will tend to be bluer 
than observed in clusters, and the inclusion of these false clusters
in the study of the Butcher-Oemler effect will tend to increase the number 
of structures with high fraction of blue galaxies.
It is therefore important to be very careful in interpreting results from 
samples which might be biased, and the need for a well defined, 
statistically significant sample of clusters to study evolutionary effects
is evident.

Garilli et al. (1995, 1996) observed clusters in the redshift range 
$0.05 \le z \le 0.25$ and used the color-magnitude relation to study 
their galaxy populations. The authors estimated the fraction of blue 
galaxies and did not find signs of evolution, but noticed
the presence of a rather large number of red galaxies, with both $g-r$
and $r-i$ colors at least $0.3^m$ redder than the early-type sequence
in the color-magnitude diagrams. These objects accounted
for $\sim 7\%$ of the total galaxy population. They also noticed that
about $1/3$ of these red galaxies had $19.5< m_r <22$ and $r-i>1$
which are typical of field galaxies at $z>0.7$, but their $g-r\sim 1.2$
colors were not red enough for these galaxies to be at such a high
redshift. The existence of such a population of galaxies would
be very important for the models of stellar population in elliptical
galaxies, and for this reason we used the same methodology as Garilli 
et al. (1996) to search for a red galaxy population in our sample. 

The slope, intercept, and scatter of the color-magnitude relation can also be 
used to study the evolution of galaxy population in clusters. Many authors 
(Rakos and Schombert 1995, Ellis et al. 1997, Gladders et al. 1998, 
Kodama et al.1998, Standford et al. 1998, Pahre 1998, 1999) have found that 
the observations are consistent with models in which most early-type galaxies 
in rich clusters are old, passively evolving systems. Bower et al. (1998) 
proposed a model on which star formation occurs over an extended period of 
time in most galaxies with star formation being truncated randomly. This type 
of star formation allows both for the small scatter of the color-magnitude 
relation which is observed up to $z \sim 1$, and for the presence of the blue 
galaxy population at intermediate redshift clusters (Butcher-Oemler effect).

Bower et al. (1998) have also shown how is it possible to use the 
color-magnitude relation as a constraint on the formation of rich clusters. 
Mergers will tend to reduce the slope and increase the scatter of the 
color-magnitude relation, therefore, the ratio between the scatter and the 
slope of the relation can be used to study the degree of merging between 
pre-existing stellar systems. The authors analyze the cases of random and 
hierarchical mergers and show that the first case would tend to very rapidly 
destroy the relation and that the second case allows for the color-magnitude 
relation to persist through a larger number of mergers.    

In this work we present the color-magnitude relation for 48 clusters, and then 
use them to study the Butcher-Oemler effect, and to search for a 
population of red galaxies in the sample. 
In order to study the Butcher-Oemler effect, we need to know the redshifts of the 
clusters. We use the 21 clusters for which there are spectroscopic redshift 
measurements to construct an empirical relation that allows the estimation of 
photometric redshifts for the remainder. 
The paper is organized as follows. In \S 2 we describe the observations 
and data reduction. \S 3 gives a description of the galaxy catalog construction. 
The results are presented in \S 4, and the summary and conclusions are  
shown in \S 5.  We will examine the evolution
of the color-magnitude relation and investigate the 
merger history of the galaxy population in a forthcoming paper.

\section{OBSERVATIONS AND DATA REDUCTION}

\subsection{Observations}

We obtained CCD images of 48 clusters in four runs 
between March 1997 and November 1998.  All data were taken with the
Tek 2K-3 detector at the Cassegrain focus of the 0.9m telescope
at the Cerro Tololo Interamerican Observatory (CTIO). The CCD 
has 2048 x 2048 pixels, each pixel covering $0.396^{\prime\prime}$, 
corresponding to a field of $\sim$ $13.5^{\prime}$ x $13.5^{\prime}$,
or $\sim 0.5-6.7$ Mpc at $z=0.03-0.38$ ($H_0 = 67$ Km s$^{-1}$ Mpc$^{-1}$).
We observed 44 Abell clusters and 4 cluster candidates detected in 
POSS-II (Second Palomar Sky Survey) (Gal et al. 1999) photographic 
plates. The images were taken through the {\it g}, {\it r} and {\it i} 
filters of the Thuan \& Gunn (1976) photometric system, with effective 
wavelengths and widths of 5118 {\AA} and 900 {\AA} 
for {\it g}, 6798 {\AA} and 1000 {\AA} for {\it r}, and 8100 {\AA} 
and 1500 {\AA} for {\it i}. The exposure times varied between 15-60 
minutes in the {\it g} band, 10-20 minutes in the
{\it r} band, and 10-15 minutes in the {\it i} band. 

Details of the observations are shown in Table 1. 
In column 1 the ACO (Abell, Corwin \& Olowin 1989) cluster number or 
the candidate identification is given; column 2 shows the observation 
date; columns 3, 4 and 5 list the exposure times in the {\it g}, 
{\it r} and {\it i} bands; the seeing is given in columns 6, 7 and 8, 
and the limiting magnitudes (see section 3.2) are listed in columns 9, 
10 and 11 for each band.
 
\subsection{Data Reduction}

The raw images were corrected for the usual instrumental effects of 
{\it ``bias''}, {\it ``dome flat''}, {\it ``sky flat''}, and 
{\it ``illumination''}. It was also necessary to apply a {\it ``shutter''} 
correction for short exposure images ($T_{exp} < 20 sec$) such as 
{\it ``dome flats''} and standard stars. In this case, the time for
the shutter to open and to close is important compared to the total 
exposure time and therefore the borders of the CCD collect photons
for a shorter time than its central part.

\subsection{Photometric Calibration}

The photometric calibration of the magnitude scale was determined
by the observation of $\sim 5$ standard stars from the Thuan \& 
Gunn (1976) list in each night. The color equation, which establishes 
the relation between instrumental and calibrated magnitudes, is 
given by eq.(1). 
\begin{equation}
g_{cal}=g_{inst}+A+B{\sec (z)}+C(g-r)_{inst}                
\end{equation}
where A is the zero point of the magnitude scale; B is the extinction 
coefficient; and C is the color term coefficient. Similar color 
equations were determined for the {\it r} and {\it i} bands, with
the color terms established from the $g-r$ and $g-i$ colors 
respectively.

When possible, the color equations were determined for each night
in the three photometric bands. Because the number of standard
stars with {\it i} calibration is very limited, there were nights
when we did not observe enough standard stars in this band to allow
us to determine this calibration. However, we found that the color 
equation coefficients were always in close agreement for the nights 
within a specific run. We therefore used the stars observed in each 
run to derive mean {\it g}, {\it r} and {\it i} color equations 
(i.e. A, B and C coefficients) for each run. The color 
equations derived for each run showed rms errors of at most 0.025$^m$
in {\it g}, 0.017$^m$ in {\it r}, and 0.009$^m$ in {\it i}. 
These errors could justify the use of the same color equations for 
each entire run, but in order to avoid small systematic errors we 
decided to use the zero points of the magnitude scale estimated 
individually for each night. The zero points determined using the mean 
extinction and color term coefficients previously calculated for each 
run are listed in Table 2, and have errors that are 
consistent with the rms values of the color equations. In Figure
1 we show the comparison between magnitudes of standard stars 
published by Thuan \& Gunn (1976) and our final calibrated measurements.

\subsubsection{External Errors}

The external photometric errors were estimated by comparison 
between the measurements of clusters that were observed in 
different nights and runs.  
Abell cluster 2700 was observed in three different
nights, and two other clusters (A248, and A324) were 
observed twice. Figure 2 shows the residuals of the magnitudes, 
in each band, for Abell 2700. The residuals are calculated from 
the galaxies brighter than the limiting magnitude (see \S 3.2) 
of the CCD images, and we found mean values of  
0.042 in {\it g}, 0.017 in {\it r}, and 0.048 in {\it i}, which are 
of the same order as the rms values of the color equation.
   
\section{GALAXY CATALOG CONSTRUCTION}

\subsection{Detection and Classification}

The detection and classification of objects in the CCD images
were performed using a modified version of the FOCAS (Faint 
Object Classification and Analysis System, Valdes 1982) package.
The first step in the construction of those catalogs is the 
detection of objects. We used a detection limit of two times 
the standard deviation of the background sky, which corresponds 
to $\sim$7\% of the local sky value. We used also a minimal 
detection area of 25 pixels, which is $\sim 2$ times greater than the 
average seeing disc.

FOCAS calculates about 30 photometric attributes for each detected 
object, including the total and aperture magnitudes. The total 
magnitude is determined with the use of an algorithm which extends
the detection isophote to lower surface brightness limits.
When the area of the extended isophote become two times bigger than
the area of the initial detection isophote the total magnitude is
computed. The aperture magnitude is calculated from the flux inside 
a central circular region of radius $5.15^{\prime\prime}$. 

Another important step in the construction of the CCD catalogs is 
the  determination of the point spread function (PSF)
which is used in the classification procedure. Stellar, non-saturated, 
objects with magnitudes between $16^m$ and $19^m$ are selected to 
construct the PSF. A visual inspection on the objects selected by
FOCAS as stars allowed us to exclude asymmetric objects, and the 
final PSF for each image was determined with typically 35 stars.
The classification of the objects was performed by FOCAS using 
the determined PSF (Valdes 1982). To avoid the inclusion of clearly 
misclassified objects we carried out a visual inspection before the 
construction of the catalog (the data are available upon request).

\subsection{Limiting Magnitude}

The histogram of total magnitudes of all the detected objects (stars and
galaxies) was used to estimate the limiting magnitude for each CCD
image. This limiting magnitude was determined by the last bin of magnitude
before the counts start to drop significantly (Picard 1991). The
values are listed in Table 1 for each cluster, in the three photometric 
bands.

\section{RESULTS}

\subsection{Color-Magnitude Diagrams}

In Figure 3 we present the $r-i \times r$ and $g-r \times r$ color-magnitude 
relations for all the clusters in our sample. The linear fit shown by a solid 
line was, 
in general, determined using the galaxies in the magnitude range between 
$M^*-1$ and $M^*+2$, where $M^*=-20.91$ ($h=0.75$) (Lin et al. 1996). 
Objects with $r<16.0$ can be saturated in our images and were excluded from 
our analysis, and for this reason the limiting magnitude to fit the 
color-magnitude relation of the lower redshift clusters was extended to 
$M^*+3$ (Abell 119, 168, 509, 1134, 2103) and $M^*+4$ (Abell 189, 261, 1260, P861C1). 
Also, because the average limiting magnitude of our sample is $r \sim 22.0$, 
we excluded fainter objects from our fitting procedure. The data was binned 
in intervals of 0.5 mag, except for the first bin for which we impose that 
there should be at least 5 galaxies. We then used ROSTAT (Beers et al. 1990) 
to determine the median color and magnitude in each bin, as well as its 
bi-weight dispersion. The fitting procedure was done with the 
GaussFit program (Jefferys et al. 1988), which allows the usage of several 
methods for solving least squares and robust estimation problems. 
After testing many of these procedures, we decided to use the ``Orthogonal 
Regression M-Estimates'' (orm), considering the color dispersion in each 
magnitude bin, to fit the color-magnitude relation.

\subsection{Photometric Redshifts}

In order to study the Butcher and Oemler effect we need to know the 
redshift of the clusters. The 21 clusters with spectroscopic measured 
redshift were used in the construction of an empirical relation between
redshift and mean $gri$ magnitudes, that was then applied 
to estimate photometric redshifts for the remainder. The idea of estimating 
cluster redshifts using magnitudes and colors is based on the fact that 
in the central region of clusters the population is dominated by early-type 
galaxies (Dressler 1980, 1997). If we assume that the same SED is typical of 
all clusters, the apparent magnitude and consequentially the colors 
of these galaxies can be used to estimate redshifts because of the strong 
4000 \AA  break feature typical of this population. We used the thirty 
brightest galaxies, after a statistical background correction, 
to estimate mean magnitudes representative of each cluster and fit
a first order polynomial relation $z(g,r,i)$.

The background correction is very important if one is trying to study 
the cluster galaxy population, and it is also a difficult matter due 
to the field-to-field fluctuations on the number counts. Usually the 
best approach in estimating the back and foreground contamination is the 
use of an annulus around the cluster. Unfortunately the 
CCD images presented in this work sample only the central region of clusters 
and therefore it is not feasible to determine corrections for each cluster 
individually. We used the mean counts in 5 control fields to estimate 
the background magnitude distribution. The mean number of interlopers  
subtracted from each field is 12 with an rms variation of $\pm 4.6$
among the 5 different control regions.

The comparison between spectroscopic and photometric redshifts is shown 
in Figure 4 and the dispersion of $z_{spect}-z_{phot}$ is 
$\sigma_{z} = 0.03$. 
This residual is of the same order as found by other works such as 
Connolly et al. (1995), Brunner et al. (1997), and Yee et al. (1999). 
Connolly et al. (1995) were the first to use an empirical approach to 
estimate photometric redshifts and using photographic data found a 
dispersion of $\sigma_{z} = 0.047$ when comparing spectroscopic and 
photometric redshifts for galaxies out to $z \sim 0.5$. Brunner et al. 
(1997), using CCD observations in 4 photometric bands, were able to obtain
a relation with dispersion of $\sigma_{z} = 0.023$ for clusters in the 
redshift range between $0.0$ and $0.4$. Recently, Yee et al. (1999) 
used the C-M relation to determine redshifts with a mean dispersion of 
$\sigma_{z} = 0.028$ for clusters at $0.1<z<0.7$.

We also estimated photometric redshifts using only the early-type 
population of the clusters. The selection of cluster early-type 
galaxies can be done with an efficiency better than $90\%$, based 
on two-color and morphological information (Pahre 1998, 1999). 
The dispersion between spectroscopic and photometric redshifts obtained 
from the early-type galaxies brighter then $21^m$ is $\sigma_{z} = 0.04$. 
Comparing this result with the previous one, it is clear that using all the 
galaxies in the central region of the clusters is consistent with using elliptical
galaxies only. This is consistent with the central region of rich, regular 
clusters being populated mainly by early-type galaxies (morphology-density 
relation, Dressler 1980, 1997).

Cluster coordinates and redshifts are presented in Table 3. 
The ACO cluster number or candidate identification is given in column 1; 
R.A. and Dec. coordinates for 1950 are listed in columns 2 and 3; the Abell
richness class is given in column 4; The redshift is given in column 5,
and the reference for this redshift is listed in column 6. Column 7 
gives the fraction of blue galaxies calculated as described in 4.3.

\subsection{Butcher-Oemler Effect}

Butcher \& Oemler (1984) (hereafter BO84) searched for signs of evolution 
in the galaxy population of clusters studying a sample of 33 clusters of 
galaxies with redshifts between 0.003 (Virgo) and 0.54. The galaxies brighter 
than $M_V=-20$, inside the central region containing 30\% of the cluster
population ($R_{30}$) were selected, and those with rest-frame $B-V$ colors at 
least 0.2 mag bluer than the ridge of the early-type galaxies in the 
color-magnitude diagrams were classified as ``blue''. The authors
defined the fraction of blue galaxies ($f_B$) as the ratio between the 
number of blue and total number of galaxies, and found that $f_B$ is
approximately constant until $z\sim0.1$ and starts to increase linearly 
for higher redshift clusters.

Our CCD images are $\sim$ $13.5^{\prime}$ x $13.5^{\prime}$ and sample 
only the central part of the clusters. For this reason it was not possible 
to determine $R_{30}$ for each cluster individually and we decided to 
adopt a fixed physical size of radius $0.7$ Mpc, which corresponds to the 
mean $R_{30}$ used by BO84.
We then selected the galaxies inside this region which, after k(z) 
correction, were found to be in the magnitude range between $M^*-1$ and 
$M^*+3$, where $M^*=-20.91$ (Lin et al. 1996) assuming $h=0.75$.
The galaxies with $g-r$ colors at least 0.2 mag bluer than the locus 
of the early-type galaxies in the color-magnitude diagrams (Figure 3) 
were identified as blue, because
a change of 0.2 mag in the $B-V$ color of an elliptical galaxy corresponds 
to approximately the same change in its $g-r$ color (Jorgensen 1995).

The background correction is very important in this study, and it tends 
to decrease $f_B$ since the field galaxies are in general bluer than the 
early-type population typically found in the central region of clusters
(Dressler 1980). BO84 estimated background corrections in different ways 
for different clusters, but their main idea was to use the population
of galaxies around the clusters. Our fields are too small to allow such 
corrections, and therefore we used 5 control fields observed during the first 
run to estimate the background correction. Since the number of background 
galaxies contaminating each cluster depends on its redshift, and the 
number of background galaxies which would be identified as blue depends 
also on each cluster C-M relation, the background correction was applied 
individually. Usually, the number of galaxies r
between $M^*-1$ and $M^*+3$ (where $M^*=-20.91$ assuming $h=0.75$, 
Lin et al. 1996) detected in a cluster image is 94, and the background 
counts are 38, with a variation of $\pm 8$ galaxies among the 5 different 
control regions. The number of blue galaxies in on average 23, and 
there are usually $9 \pm 2$ background galaxies that would be selected as blue. 

Several corrections need to be applied to guarantee that the same physical
region, and the same interval in the luminosity function (LF) of the clusters, 
are being used to compute the fraction of blue galaxies. Clusters at 
low redshift suffer from two different problems: 1) our CCD fields are not 
big enough to observe a region of radius $0.7$ Mpc; and 2) the brightest 
objects (around $M^*-1$) are saturated. For clusters at higher redshifts, 
the fainter objects (around $M^*+3$) are not detected. To estimate the 
effect of measuring $f_B$ at different regions in the cluster we used 5
clusters in the redshift range $z=0.16-0.18$ for which the angular sizes 
are small, and for which the entire magnitude interval from $M^*-1$ to 
$M^*+3$ can be observed. The mean fraction of blue galaxies computed at 
different radius is shown in Figure 5, and the errorbars are $rms/\sqrt{5}$. 
Fitting a linear relation to the points we find: 
\begin{equation}
f_B=0.103 \times(R/0.7)+0.185 \hskip 1truecm   (\sigma_{f_B}=0.093) 
\end{equation}
where $R$ is the radius in Mpc.

The idea that the fraction of blue galaxies increases in the outer parts 
of the cluster is in agreement with previous results. This behavior was 
already noticed by BO84 in their original work, and Rakos et al. (1997) 
also observe this trend in the detailed study of Abell 2317. This is 
also consistent with the work by Dressler and Gunn (1992), Dressler et al. 
(1994) and Oemler et al. (1997), which shows 
that early-type galaxies cluster more strongly than those with signs of 
recent or ongoing star-formation.

To estimate the effect of losing objects off the bright or faint ends of
the chosen magnitude range, we use the sample of clusters at moderate 
redshift ($0.085<z<0.175$), for which we are able to observe the entire 
interval of $M^*-1$ to $M^*+3$ and therefore to calculate $f_B$ at various 
limiting magnitudes. 
The sample was subdivided in two ($0.085<z<0.130$ and $0.130<z<0.175$) 
in order to check if clusters at different redshifts show a different 
distribution of luminosities of its blue galaxy population. Figure 6a shows 
the result of missing the cluster brightest objects ($f_B$ versus $\Delta_{mag}$, 
where  $\Delta_{mag} = (M^*+3)-(M^*-?)$), and Figure 6b indicates the effect of 
losing the faintest ones ($f_B$ versus $\Delta_{mag}$, where 
$\Delta_{mag} = (M^*+?)-(M^*-1)$). Two main results can be draw from these 
figures: (1) the fraction of blue galaxies inside a fixed radius of $0.7$ Mpc 
is larger for clusters at higher redshift (open circles) than for the lower 
redshift ones (open squares) (Buther-Oemler effect); 
(2) $f_B$ increases as brighter objects are 
excluded from the sample (Fig 6a) and decreases if the faint objects are 
missing (Fig 6b), indicating that the blue galaxy population is faint. The 
results are the same for the two subsamples, and the slope of the $f_B$ versus 
$\Delta_{mag}$ relations does not show significant change with redshift. We 
therefore use the entire sample of clusters (between $0.085<z<0.175$) to 
estimate the effect of measuring $f_B$ at different intervals of magnitude.
The following two equations indicate the linear fits
for losses from the bright and faint ends, respectively, obtained from 
the total mean points (solid dots). 
\begin{equation}
f_B= -0.034 \times \Delta_{mag}+0.362 \hskip 1truecm (\sigma_{f_B}=0.064)  
\end{equation}
\begin{equation}
f_B= +0.055 \times \Delta_{mag}+0.016  \hskip 1truecm (\sigma_{f_B}=0.050)   
\end{equation}
The point at $\Delta_{mag}=2.0$ in Figure 6b, corresponding to the 
magnitude interval between $M^*-1$ and $M^*+1$, was excluded from  
the fitting procedure. This point was ignored because when the faint 
objects are excluded the number of cluster galaxies drops dramatically, 
significantly increasing the errors in $f_B$. This effect is not as important 
when the brightest objects are missing, since the number of faint objects 
is always much larger than the number of bright objects. Although BO84
were limited to much brighter objects, they find the same tendency 
observed in our sample. The blue galaxies are also the faintest ones.
Rakos et al. (1997) found an opposite trend in A2317 ($z=0.211$). The authors 
calculated $f_B$ at four magnitude bins and found that $f_B$ is highest in the 
brighter bin, drops in the two intermediate bins, and rises again in the fainter
one. The fact that the blue galaxies in more evolved, low redshift clusters are 
faint, and that a greater number of bright blue galaxies is found in higher 
redshift or less evolved clusters seems to support the idea that BO galaxies 
are late-type, low surface brightness objects, which fade after a burst of star 
formation (RS95).  
 
The CCD images presented here are deeper than the photographic plates 
used by BO84, therefore allowing us to include fainter galaxies in our 
$f_B$ estimate. On the other hand, BO84 were able to measure the brightest 
cluster objects that are usually saturated in our images. As it can be 
seen from Figures 6a and 6b the differences in the magnitude interval used 
to determine $f_B$ strongly affects its measurements. In order to compare our 
results with BO84 we need to ``correct'' our fraction of blue galaxies as if we 
had used the same limiting magnitudes. If we assume that the 
typical $g-r$ color of an elliptical galaxy at $z=0$ is 0.44 (Small 
1996), the $M_V=-20$ limiting magnitude used by BO84 corresponds 
approximately to $M_r=-20.19$ (Jorgensen et al. 1995). This is the same as 
limiting our sample at $M^*+0.72$, which results in a reduction of 0.13 
(Figure 6b) from our $f_B$ values. BO84 give no
limiting magnitude for brighter objects, so we 
apply no correction for this effect. However, we do not expect this to be a 
major problem since it can be seen from Figures 6a and 6b (and equations 3 
and 4) that the effect of missing bright objects is weaker than 
that of missing faint objects.  Comparing the two clusters 
that we have in common with BO84, we find that their $f_B$ measurements 
are $0.13$ (for Abell 1689) and $0.16$ (for Abell 370)
lower than ours, indicating that our correction of 
0.13 is reasonable.
 
The final fractions of blue galaxies are shown in column 7 of Table 3 and 
in Figure 7a. The quoted $f_B$ errors are $(N)^{1/2}$ of the number of 
galaxies after background correction. Abell 1993 shows as extremely high 
fraction of blue galaxies ($0.606 \pm 0.141$) that can be due to a wrong 
redshift estimation and/or to the common problem of projection effects
existent in catalogs of clusters constructed from visual inspection of 
images. This is a not very well studied cluster and the last reference 
is from Abell et al. (1989) and for this reason we decide not to include 
it in Figure 7 and in the following discussion. Figure 7b shows the mean $f_B$ 
values for intervals of 0.04 in redshift for the clusters with spectroscopic 
redshifts (filled circles), and for the clusters with estimated photometric 
redshifts (open circles). Both photometric and spectroscopic redshift
samples show the same behavior in the $f_B \times z$ plot. 
In Figure 7c we plot the mean $f_B$ values for the clusters in our sample  
(filled circles), for BO84's clusters (open squares), and for  
10 x-ray clusters observed by Smail et al. (1998) (crosses). 
The solid and dashed lines shown in figure 7 represent respectively 
the relation originally found by BO84, and the one estimated by 
Rakos and Schombert (1995, hereafter RS95) in their study of 17 clusters.
The dotted line represent a linear fit to the binned mean points from our 
sample.  

In order to check the adherence of the points to BO84, RS95 and to our 
estimated $f_B \times z$ relation we plot in Figure 7d the accumulated 
product $f_B \times \Delta z$ for our data and for the three relations. 
The data seems more consistent with our relation and also with RS95 
relation which is very similar to ours except for an offset due to the 
negative fraction of blue galaxies that we find at lower redshifts. The 
data shows an stronger increase in the fraction of blue galaxies with 
redshift than estimated by B084. The dot-dash line indicates the behavior 
assuming no evolution, or a constant $f_B$ of 0.09. 
This is a mean value estimated from the clusters with spectroscopic 
redshiftin our sample, and it is clear that $f_B$ is not constant: there 
is an absence of blue galaxies at local clusters, and an excess of blue 
objects at higher redshifts. 

Andreon and Ettori (1999) used x-ray measurements of BO84 sample to argue that 
these clusters are not part of the same class of objects, and that therefore 
the increase in the fraction of blue galaxies with redshift might be simply a 
selection effect. Smail et al. (1998) studied the Butcher-Oemler effect in 
clusters at $0.22-0.28$ with similar x-ray luminosity, which should result in 
a sample consisting of the same kind of objects, and did not find the evolutionary 
trend shown in BO84, RS95 and in our data. Their observations, binned in 
redshift, are indicated by two crosses in Figure 7c and lie well below previous 
observations. It is interesting to note that if we concentrate our attention only 
in the redshift range between $0.08$ and $0.2$ (see Figure 7a), where we 
observed 10 times more clusters than BO84 and where most of our sample lies, 
we find objects with a wide range in $f_B$. These observations, as well as Smail 
et al. (1998) measurements, might be indicating that the Butcher-Oemler effect 
is at least partially due to selection effects that are still not well 
understood. A complete, statistically significant sample of clusters, 
representative of all kinds of clusters at a certain redshift, should 
be able to resolve this question. Unfortunately, such a complete sample is very 
hard to observe, and any sample will always be biased toward richer, 
more luminous systems, but a very large sample should contain most kinds of 
clusters at a given redshift. 
With this purpose we are now analyzing a sample of $\sim 500$ clusters that 
were observed in the same way described in this work. This large CCD sample of 
Abell clusters was primarily obtained to calibrate the POSS-II (Second Palomar 
Sky Survey) 
photographic plates and is now also being used to study the population of 
galaxies in clusters and its evolution.
  
\subsection{Anomalously Red Galaxies?}

In each cluster we selected as red all the galaxies with $g-r$ and $r-i$ colors 
$0.3^{m}$ above the lines in Figure 3, and brighter than $m_r = 22^{m}$. The upper 
panels in Figure 8 show the histograms of $m_r$, $g-r$, and $r-i$ of the red 
galaxies. The filled histograms represent objects with $r-i \geq 1$.
As it can be seen from upper panel, most of those objects are faint having $m_r$ 
between $20.5^m$ and $22^m$ and have $g-r$ colors varying from $\sim 0.8$ up to 
$2.4$. Objects with $g-r\sim 1.8$ are red enough to be in agreement with the 
observation of a high redshift ($z \sim 0.7$) population of galaxies with 
$r-i \geq 1$. Such red colors are due to the heavy k-corrections for elliptical 
galaxies at $z>0.4$ (see Figure 2 of Gal et al. 1999 for $(g-r)\times(r-i)$ 
tracks for diferent types of galaxies as a function of redshift). At $z\sim 0.4$ 
the 4000 \AA  break feature typical of elliptical galaxies enters the $g$ band 
causing a dramatic drop in its blue luminosity. The objects with $g-r\sim 1.2$ 
seem to correspond to the anomalously red galaxies found by Garilli.

Most of the stars with $r-i \geq 1$ have $g-r\sim 1.2$ (Kennefick 1996), so 
stellar contamination could be responsible for at least part of this red 
population. In order to study the seeing effect in this population, the sample 
was divided in two: one with better seeing ($FWHM< 1.44^{\prime\prime}$), and 
other with worse seeing ($FWHM> 1.44^{\prime\prime}$) in {\it r} band. The bad 
seeing observations shown in the central panels of Figure 8 present mainly one 
peak around $1.2^m$ on the $g-r$ distribution. But the $g-r$ distribution
from the better seeing images (lower panels in Figure 8) shows also the 
presence of redder objects, which
indicates that at least part of those anomalously red objects are in fact stars 
that were misclassified as galaxies. This effect should be even stronger for the
Garilli et at. (1996) data, which had an average poor seeing of $2^{\prime\prime}$.
Furthermore, only $42\%$ of the ``anomalously'' red galaxies
classified as galaxies in $r$ were also classified as galaxies in $g$
and $i$, while $71\%$ of the rest of the $r$ galaxies were classified
as galaxies in all filters.

\section{SUMMARY AND CONCLUSIONS}

We present a photometric study of 48 clusters of galaxies. Color-magnitude
diagrams are shown for the entire sample, and used to study the 
Butcher-Oemler effect and the population of red galaxies.
We also present photometric redshifts for the clusters for which there 
are not spectroscopic measurements. The photometric redshift is 
estimated from the colors and magnitudes of cluster galaxies and has an 
accuracy of $\sim 0.03$. Our main conclusions are summarized as follows:

\begin{enumerate}
 
\item The blue, Butcher-Oemler galaxies, lie preferentially in the 
outer edges of the cluster, and are in general fainter than the cluster 
early-type population.

\item We find that the increase in the fraction of blue galaxies 
with redshift seems more consistent with the relation estimated by RS95 
than with the original one by BO84. We predict a linear $f_B \times z$ 
relation consistent with RS95 estimation for $z>0.08$ but in which the
fraction of blue galaxies continues to decrease for lower redshifts. 
This result is consistent with the observation of a smaller number 
of blue galaxies in the cluster region than would be observed in a 
blank field region of the same angular area. 
However, there is no redshift evolution of $f_B$ in the
range between $\sim 0.08$ and $0.2$, where most of our clusters lie.
Also, a large range of $f_B$ values is seen between $\sim 0.2$ and $0.3$ 
when Smail et al. (1998) x-ray clusters are added to the $f_B \times z$ 
plot. We are currently analyzing a sample of $\sim 500$ clusters which 
might help in understanding the selection effects and may clarify the results.    

\item Studying the population of red galaxies in our images
we found that some of these objects have colors that are normal for
background galaxies, and that the colors of the supposably anomalous red
objects are typical of stars. The fact that these objects are near the
limiting magnitude of our sample, and that they are found in greater
number at the bad seeing images, lead us to the conclusion that at
least part of those objects are stars that were misclassified as galaxies.

\end{enumerate}

\acknowledgments

We would like to thank D. Wittman, J.A. Tyson, H.V. Capelato, R.R. Gal,
and S.C. Odewhan for very helpful comments and suggestions which helped
to improve the paper. We also thank the anonymous referee for the detailed 
revision and usefull recommendations provided for this work.


\clearpage
\begin{deluxetable}{ccccccccccc}
\footnotesize \tablecaption{Log of the Observations
\label{tbl-1}} \tablewidth{0pt} \tablehead{
\colhead{ACO} &\colhead{Date} &\multicolumn {3}{c}{$T_{exp}$(s)} &\multicolumn {3}{c}
{seeing(")} &\multicolumn {3}{c}{Lim Mag} \nl
\colhead{} &\colhead{} &\colhead{\it g} &\colhead{\it r} &\colhead{\it i} &\colhead{\it g} &\colhead{
\it r} &\colhead{\it i} &\colhead{\it g} &\colhead{\it r} &\colhead{\it i} \nl}
\startdata 
1134& 97, March 08 & 1800 & 1200& 900 & 1.60 & 1.45 & 1.30 & 22.7 & 22.3 & 21.9\nl
P861C1&97, March 08& 1200 & 900 & 900 & 1.79 & 1.74 & 1.42 & 22.5 & 22.5 & 21.7\nl
820 & 97, March 09 &  900 & 600 & 600 & 1.40 & 1.44 & 1.16 & 22.3 & 22.1 & 21.3\nl
827 & 97, March 09 &  900 & 600 & 600 & 1.58 & 1.45 & 1.48 & 22.5 & 22.3 & 21.7\nl
860 & 97, March 09 &  900 & 600 & 600 & 1.47 & 1.42 & 1.37 & 22.3 & 22.1 & 21.3\nl
1191& 97, March 09 &  900 & 600 & 600 & 1.58 & 1.38 & 1.33 & 22.5 & 22.1 & 21.5\nl
1577& 97, March 09 & 1800 & 900 & 900 & 1.39 & 1.29 & 1.27 & 22.9 & 22.5 & 21.9\nl
2066& 97, March 09 & 1800 & 900 & 900 & 1.49 & 1.26 & 1.24 & 22.9 & 22.5 & 21.9\nl
810 & 97, March 10 & 1800 & 900 & 900 & 1.34 & 1.41 & 1.26 & 22.9 & 22.3 & 21.7\nl
1650& 97, March 10 &  900 & 600 & 600 & 1.74 & 1.68 & 1.53 & 22.5 & 22.1 & 21.7\nl
1689& 97, March 10 &  900 & 600 & 600 & 1.32 & 1.20 & 1.19 & 22.5 & 22.3 & 21.7\nl
2094& 97, March 10 & 1800 & 900 & 900 & 1.21 & 1.27 & 1.38 & 22.7 & 22.3 & 21.7\nl
2103& 97, March 10 & 1800 & 900 & 900 & 1.34 & 1.37 & 1.32 & 22.7 & 22.3 & 21.7\nl
933 & 97, March 11 & 1800 & 900 & 900 & 1.59 & 1.37 & 1.48 & 22.9 & 22.3 & 21.7\nl
944 & 97, March 11 &  900 & 600 & 600 & 1.41 & 1.29 & 1.21 & 22.5 & 22.3 & 21.3\nl
1620& 97, March 11 &  900 & 600 & 600 & 1.37 & 1.23 & 1.18 & 22.5 & 22.1 & 21.5\nl
2128& 97, March 11 &  900 & 600 & 600 & 1.27 & 1.16 & 1.22 & 22.3 & 22.1 & 21.5\nl
\tableline                                                 
248  & 97, Nov 23 & 2700 & 900 & 900 & 1.35 & 1.46 & 1.46 & 23.3 & 22.7 & 22.3\nl
508  & 97, Nov 23 & 3600 & 900 & 900 & 1.32 & 1.36 & 1.43 & 22.9 & 23.1 & 22.7\nl
2700 & 97, Nov 23 & 2700 & 900 & 900 & 1.83 & 1.52 & 1.52 & 23.5 & 22.7 & 22.1\nl
145  & 97, Nov 26 & 1800 & 900 & 900 & 1.67 & 1.41 & 1.27 & 23.1 & 22.5 & 21.9\nl
248  & 97, Nov 26 & 1800 & 900 & 900 & 1.73 & 1.46 & 1.63 & 23.3 & 22.3 & 21.9\nl
324  & 97, Nov 26 & 1800 & 900 & 900 & 1.92 & 1.75 & 1.67 & 23.1 & 22.5 & 21.5\nl
2700 & 97, Nov 26 & 1800 & 900 & 900 & 1.80 & 1.46 & 1.37 & 23.3 & 22.3 & 21.9\nl
\tableline                                                
1200 & 98, Apr 22 & 1200 & 900 & 900  & 1.89 & 1.76 & 1.55 & 22.7 & 22.3 & 21.9\nl
1260 & 98, Apr 22 & 1200 & 900 & 900  & 1.53 & 1.43 & 1.54 & 23.3 & 22.5 & 21.7\nl
1238 & 98, Apr 24 & 1200 & 900 & 900  & 1.90 & 1.53 & 1.31 & 22.7 & 22.3 & 21.7\nl
1373 & 98, Apr 24 & 1200 & 900 & 900  & 2.09 & 1.50 & 1.40 & 22.9 & 22.3 & 21.7\nl
1525 & 98, Apr 24 & 1200 & 900 & 900  & 1.37 & 1.33 & 1.31 & 22.9 & 22.5 & 21.9\nl
1882 & 98, Apr 24 & 1200 & 900 & 900  & 1.47 & 1.37 & 1.39 & 22.9 & 22.5 & 21.9\nl
1993 & 98, Apr 24 & 1200 & 900 & 900  & 1.80 & 1.86 & 2.15 & 22.7 & 22.5 & 21.7\nl
1399 & 98, Apr 26 & 1200 & 900 & 900  & 1.63 & 1.49 & 1.43 & 22.5 & 22.1 & 21.5\nl
1938 & 98, Apr 26 & 1200 & 900 & 900  & 1.65 & 1.75 & 1.60 & 22.5 & 22.1 & 21.7\nl
2051 & 98, Apr 26 & 1200 & 600 & 600  & 1.61 & 1.60 & 1.46 & 22.3 & 21.9 & 21.3\nl
2053 & 98, Apr 26 & 1200 & 900 & 900  & 1.75 & 1.67 & 1.53 & 22.7 & 22.3 & 21.5\nl
\tableline                                                
189  & 98, Nov 13 & 2700 & 900 & 900 & 1.74 & 1.35 & 1.28 & 23.5 & 22.5 & 21.9\nl
482  & 98, Nov 13 & 2700 & 900 & 900 & 1.56 & 1.70 & 1.42 & 23.3 & 22.7 & 22.1\nl
2700 & 98, Nov 13 & 2700 & 900 & 900 & 1.39 & 1.11 & 1.16 & 23.5 & 22.5 & 21.9\nl
352  & 98, Nov 14 & 2700 & 900 & 900 & 1.62 & 1.34 & 1.47 & 23.5 & 22.5 & 21.9\nl
509  & 98, Nov 14 & 2700 & 900 & 900 & 1.24 & 1.21 & 1.28 & 23.3 & 22.5 & 21.9\nl
P887C04 & 98, Nov 14 & 3600 & 900 & 900 & 1.97 & 1.56 & 1.67 & 23.1 & 22.3 & 21.5\nl
261  & 98, Nov 15 & 2700 & 900 & 900 & 1.88 & 1.39 & 1.88 & 23.3 & 22.5 & 21.9\nl
324  & 98, Nov 15 & 2700 & 900 & 900 & 1.43 & 1.30 & 1.30 & 23.3 & 22.5 & 21.9\nl
776  & 98, Nov 15 & 2700 & 900 & 900 & 1.76 & 1.71 & 1.49 & 23.3 & 22.3 & 21.7\nl
168  & 98, Nov 16 & 2700 & 900 & 900 & 1.58 & 1.21 & 1.19 & 23.5 & 22.5 & 21.9\nl
255  & 98, Nov 16 & 2700 & 900 & 900 & 1.73 & 1.49 & 1.50 & 23.5 & 22.5 & 21.9\nl
P887C05 & 98, Nov 16 & 2700 & 900 & 900 & 1.53 & 1.33 & 1.47 & 23.3 & 22.3 & 21.7\nl
119  & 98, Nov 17 & 2700 & 900 & 900 & 1.69 & 1.50 & 1.45 & 23.5 & 22.5 & 21.9\nl
370  & 98, Nov 17 & 2700 & 900 & 900 & 1.84 & 1.83 & 1.64 & 23.5 & 22.7 & 21.9\nl
477  & 98, Nov 17 & 2700 & 900 & 900 & 1.53 & 1.39 & 1.31 & 23.3 & 22.5 & 22.1\nl
P887C18 & 98, Nov 17 & 2700 & 900 & 900 & 2.02 & 1.83 & 1.81 & 23.3 & 22.3 & 21.7\nl
\tableline                                                    
\enddata
\end{deluxetable}
 

\begin{deluxetable}{lccc}
\footnotesize \tablecaption{Zero points of the magnitude scale 
\label{tbl-2}} \tablewidth{0pt} \tablehead{
\colhead{NIGHT} &\colhead{$A_g$} &\colhead{$A_r$} &\colhead{$A_i$}\nl}
\startdata
1, 97, March 08 &  -8.643 $\pm$0.036 & -8.388 $\pm$0.011 & -8.391 $\pm$0.007 \nl
2, 97, March 09 &  -8.627 $\pm$0.030 & -8.382 $\pm$0.016 & -8.379 $\pm$0.054 \nl
3, 97, March 10 &  -8.645 $\pm$0.032 & -8.392 $\pm$0.008 & -8.406 $\pm$0.005 \nl
4, 97, March 11 &  -8.596 $\pm$0.032 & -8.366 $\pm$0.009 & -8.384 $\pm$0.002 \nl
\tableline
5, 97, Nov 23 &  -8.296 $\pm$0.026 & -7.970 $\pm$0.048 & -7.694 $\pm$0.012 \nl
6, 97, Nov 26 &  -8.318 $\pm$0.008 & -8.007 $\pm$0.034 & -7.706 $\pm$0.007 \nl
\tableline
7, 98, Apr 22 & -8.163 $\pm$0.088 & -8.079 $\pm$0.018 & -8.139 $\pm$0.004 \nl
8, 98, Apr 24 & -8.141 $\pm$0.073 & -8.050 $\pm$0.041 & -8.101 $\pm$0.030 \nl
9, 98, Apr 26 & -8.223 $\pm$0.030 & -8.116 $\pm$0.025 & -8.180 $\pm$0.074 \nl
\tableline
10, 98, Nov 13 & -8.180 $\pm$0.021 & -7.966 $\pm$0.029 & -8.085 $\pm$0.003 \nl
11, 98, Nov 14 & -8.185 $\pm$0.025 & -7.978 $\pm$0.048 & -8.087 $\pm$0.001 \nl
12, 98, Nov 15 & -8.184 $\pm$0.026 & -7.966 $\pm$0.022 & -8.098 $\pm$0.019 \nl
13, 98, Nov 16 & -8.167 $\pm$0.022 & -7.944 $\pm$0.035 & -8.035 $\pm$0.022 \nl
14, 98, Nov 17 & -8.183 $\pm$0.022 & -7.965 $\pm$0.018 & -8.083 $\pm$0.003 \nl
\enddata
\end{deluxetable}
\clearpage


\clearpage

\begin{deluxetable}{ccccccc}
\footnotesize \tablecaption{Characteristics of the Sample
\label{tbl-3}} \tablewidth{0pt} \tablehead{
\colhead{ACO} &\colhead{R.A.(1950)} &\colhead{Dec.(1950)} &\colhead{R} &\colhead{$z$} 
&\colhead{Ref.($z$)} &\colhead{$f_B$} \nl}
\startdata
 119    &  00 53.8 &  -01 32 &  1 & 0.0440  &   SR91  &   -0.098  $\pm$ 0.007 \nl 
 145    &  01 04.2 &  -02 43 &  2 & 0.1909  &   F93   &    0.205  $\pm$ 0.045 \nl 
 168    &  01 12.6 &  -00  1 &  2 & 0.0452  &   SR91  &   -0.042  $\pm$ 0.021 \nl 
 189    &  01 21.1 &  +01 23 &  1 & 0.0325  &   SR91  &   -0.085  $\pm$ 0.017 \nl 
 248*   &  01 42.3 &  -02 31 &  1 & 0.184   &  pho    &    0.233  $\pm$ 0.047 \nl 
 255    &  01 44.8 &  -02 14 &  1 & 0.086   &  pho    &    0.192  $\pm$ 0.041 \nl 
 261    &  01 48.9 &  -02 29 &  1 & 0.0477  &   SR91  &   -0.076  $\pm$ 0.019 \nl 
 324*   &  02 11.2 &  -01 46 &  1 & 0.167   &  pho    &    0.108  $\pm$ 0.035 \nl 
 352    &  02 25.6 &  -02 24 &  1 & 0.170   &  pho    &    0.159  $\pm$ 0.038 \nl 
 370    &  02 37.3 &  -01 48 &  0 & 0.3750  &   SR91  &    0.243  $\pm$ 0.054 \nl 
 477    &  04 09.6 &  -02  0 &  1 & 0.193   &  pho    &    0.153  $\pm$ 0.031 \nl 
 482    &  04 14.0 &  -02 15 &  1 & 0.149   &  pho    &    0.079  $\pm$ 0.027 \nl 
 508    &  04 43.3 &  +01 55 &  2 & 0.1479  &   SR91  &    0.035  $\pm$ 0.021 \nl 
 509    &  04 45.1 &  +02 12 &  1 & 0.0836  &   SR91  &    0.024  $\pm$ 0.022 \nl 
 776    &  09 13.7 &  -00 11 &  1 & 0.202   &  pho    &    0.163  $\pm$ 0.025 \nl 
 810    &  09 28.2 &  -01 56 &  0 & 0.103   &  pho    &    0.004  $\pm$ 0.017 \nl 
 820    &  09 31.3 &  -02 42 &  1 & 0.084   &  pho    &    0.112  $\pm$ 0.025 \nl 
 827    &  09 32.1 &  -02 44 &  1 & 0.143   &  pho    &    0.213  $\pm$ 0.057 \nl 
 860    &  09 41.0 &  +02 20 &  1 & 0.179   &  pho    &    0.207  $\pm$ 0.046 \nl 
 933    &  10 05.1 &  +00 46 &  0 & 0.0951  &   SR91  &   -0.047  $\pm$ 0.012 \nl 
 944    &  10 08.6 &  -01 47 &  1 & 0.161   &  pho    &    0.108  $\pm$ 0.025 \nl 
1134    &  10 54.5 &  -01 52 &  2 & 0.108   &  pho    &    0.159  $\pm$ 0.034 \nl 
1191    &  11 08.6 &  +01  2 &  2 & 0.136   &  pho    &    0.144  $\pm$ 0.025 \nl 
1200    &  11  9.8 &  -02 53 &  1 & 0.104   &  pho    &   -0.054  $\pm$ 0.010 \nl 
1238    &  11 20.4 &  +01 22 &  1 & 0.0716  &   SR91  &   -0.024  $\pm$ 0.017 \nl 
1260    &  11 23.8 &  +02 20 &  2 & 0.0492  &  pho    &    0.026  $\pm$ 0.029 \nl 
1373    &  11 42.9 &  -02 07 &  2 & 0.1314  &   SR91  &    0.136  $\pm$ 0.031 \nl 
1399    &  11 48.6 &  -02 49 &  2 & 0.0913  &   SR91  &    0.185  $\pm$ 0.041 \nl 
1525    &  12 19.5 &  -00 52 &  3 & 0.2590  &   SR91  &    0.121  $\pm$ 0.029 \nl 
1577    &  12 35.3 &  -00  0 &  1 & 0.133   &  pho    &    0.101  $\pm$ 0.028 \nl 
1620    &  12 47.2 &  -01 19 &  0 & 0.0825  &  LO95   &    0.035  $\pm$ 0.022 \nl 
1650    &  12 56.2 &  -01 29 &  2 & 0.0845  &   SR91  &   -0.011  $\pm$ 0.012 \nl 
1689    &  13 09.0 &  -01  6 &  4 & 0.1832  &   SR91  &    0.093  $\pm$ 0.019 \nl 
1882    &  14 12.1 &  -00 06 &  3 & 0.110   &  pho    &    0.140  $\pm$ 0.026 \nl 
1938    &  14 35.2 &  -00 03 &  1 & 0.163   &  pho    &    0.086  $\pm$ 0.027 \nl 
1993    &  14 53.4 &  +02 03 &  1 & 0.245   &  pho    &    0.606  $\pm$ 0.141 \nl 
2026    &  15 06.0 &  -00 05 &  1 & 0.0874  &  SR91   &   -0.120  $\pm$ 0.002 \nl 
2051    &  15 14.2 &  -00 46 &  2 & 0.205   &  pho    &    0.233  $\pm$ 0.054 \nl 
2053    &  15 14.7 &  -00 30 &  1 & 0.1127  &   SR91  &    0.191  $\pm$ 0.040 \nl 
2066    &  15 21.4 &  +01 13 &  2 & 0.144   &  pho    &    0.087  $\pm$ 0.020 \nl 
2094    &  15 34.0 &  -01 52 &  1 & 0.1445  &  S98    &    0.245  $\pm$ 0.046 \nl 
2103    &  15 37.3 &  -02  0 &  0 & 0.101   &  pho    &    0.088  $\pm$ 0.041 \nl 
2128    &  15 46.3 &  -02 54 &  0 & 0.1019  &  P92    &    0.008  $\pm$ 0.024 \nl 
2700**  &  00 01.3 &  +01 47 &  1 & 0.0978  &  SR91   &    0.145  $\pm$ 0.043 \nl 
P861C01 &  00 23.5 &  +01 28 & \nodata & 0.065   &  pho    &    0.017  $\pm$ 0.025 \nl 
P887C04 &  06 39.1 &  +02 34 & \nodata & 0.150   &  pho    &    0.077  $\pm$ 0.030 \nl 
P887C05 &  08 34.3 &  +01 42 & \nodata & 0.229   &  pho    &    0.131  $\pm$ 0.033 \nl 
P887C18 &  09 18.2 &  +01 09 & \nodata & 0.126   &  pho    &    0.062  $\pm$ 0.029 \nl 
\enddata
\tablenotetext{*}{mean of two observations}
\tablenotetext{**}{mean of three observations}
\tablenotetext{}{SR91 - Struble \& Rood 1991}
\tablenotetext{}{F93  - Fetisova et al. 1993}
\tablenotetext{}{LO95 - Ledlow \& Owen 1995}
\tablenotetext{}{S98  - Slinglend et al. 1998}
\tablenotetext{}{P92  - Postman et al. 1992}
\end{deluxetable}

\clearpage

\begin{figure}
\figurenum{1}
\plotone{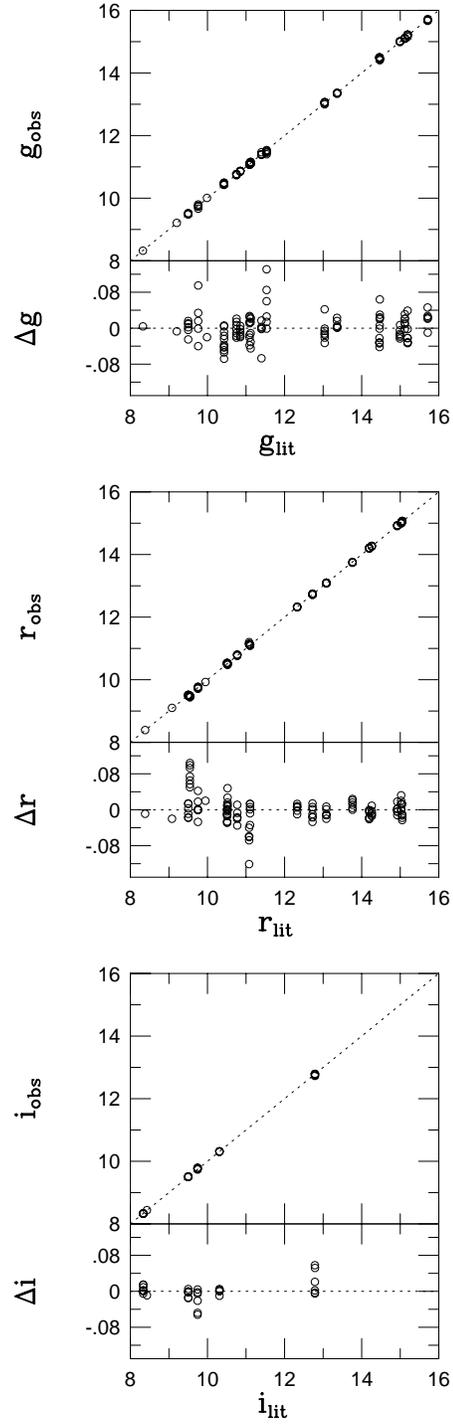}
\caption{Comparison between the magnitudes of standard
stars published by Thuan \& Gunn (1976) and our calibrated measurements.}
\end{figure}

\begin{figure}
\figurenum{2}
\plotone{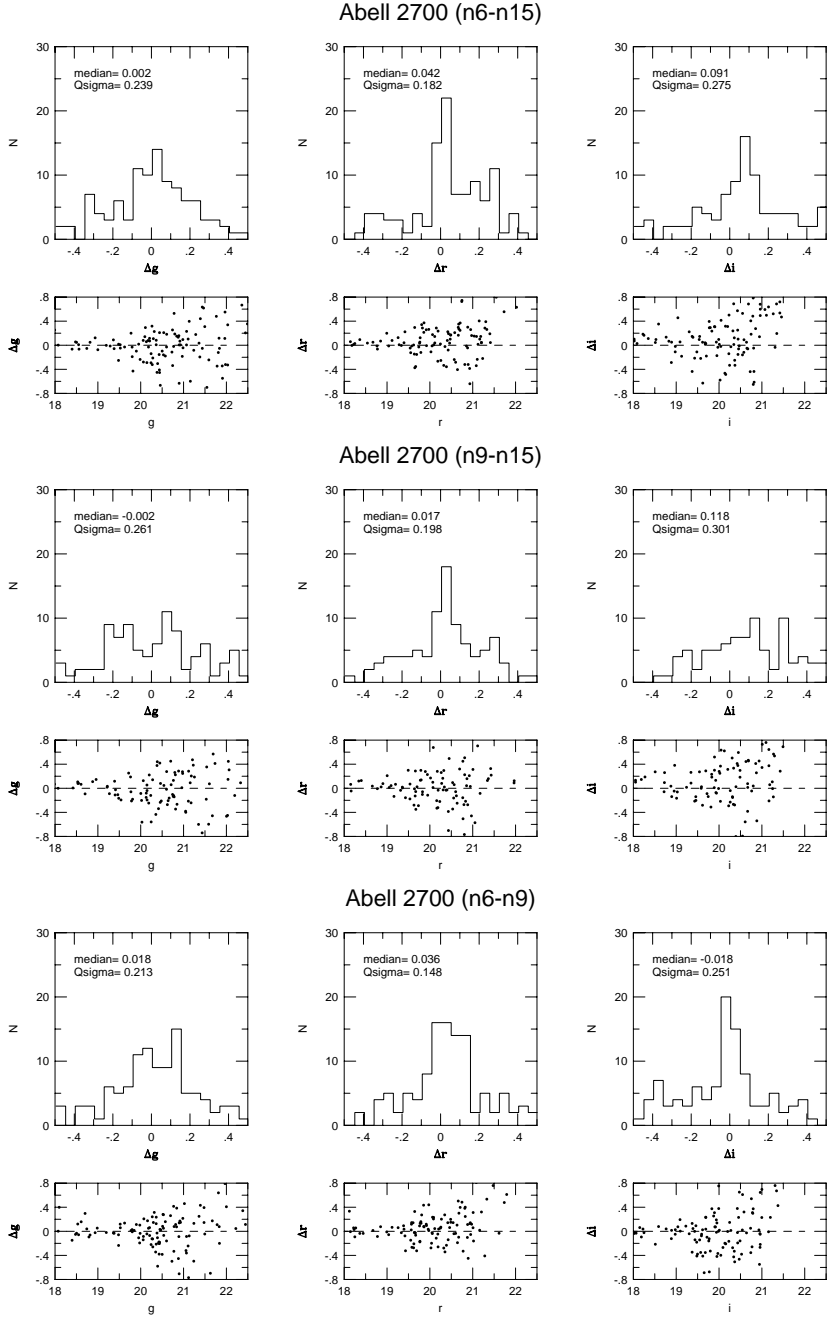}
\caption{Histogram of the magnitude residuals and its correlation
with apparent magnitude for Abell 2700. The figure show comparisons
between measurements obtained in three different nights.}
\end{figure}

\begin{figure}
\figurenum{3}
\plotone{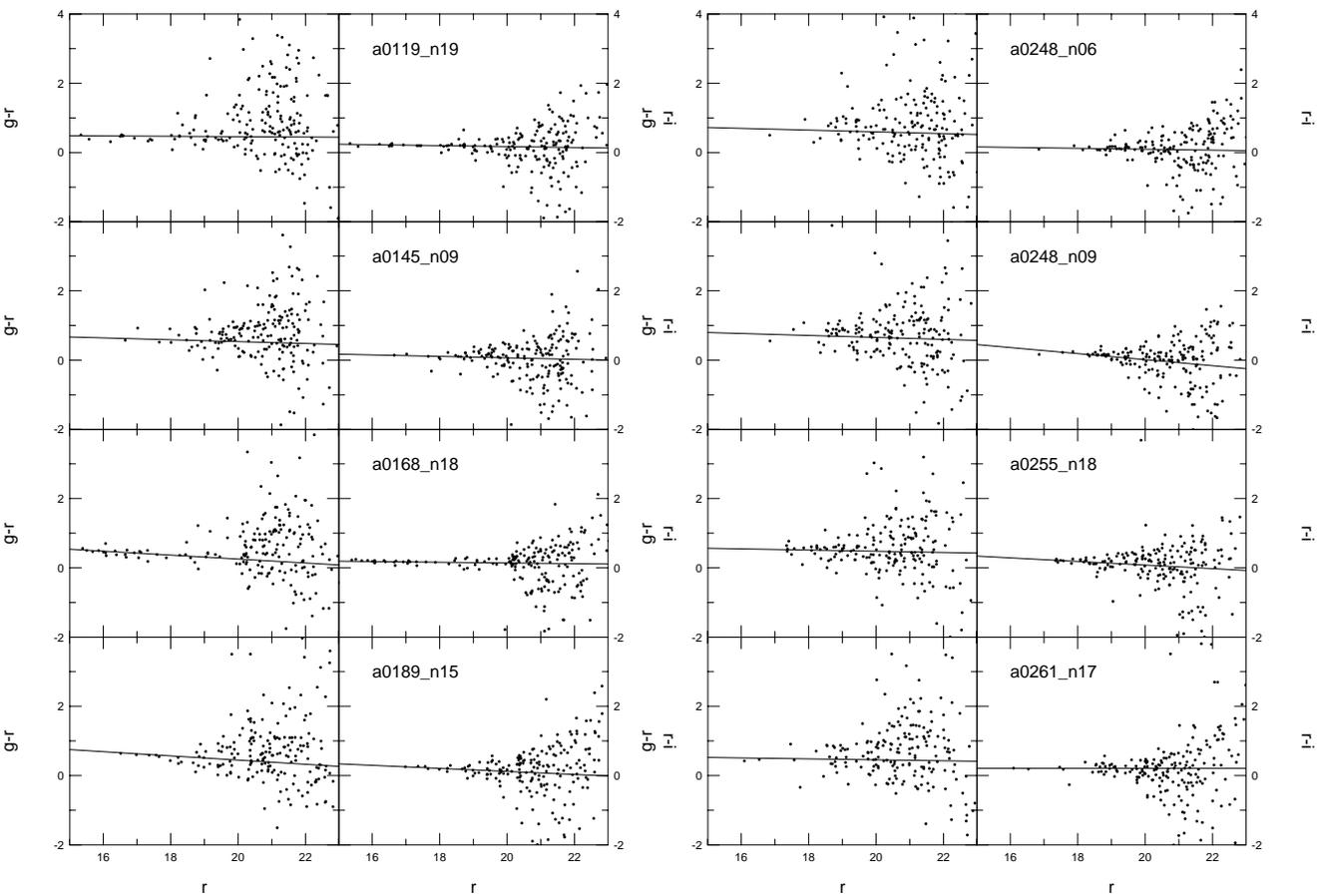}
\caption{$g-r\times r$, and $r-i\times r$ color-magnitude 
diagrams for the cluster sample. The solid lines indicates the fitting 
of a linear relation which represents the locus of early-type galaxies.}
\end{figure}

\begin{figure}
\figurenum{3}
\plotone{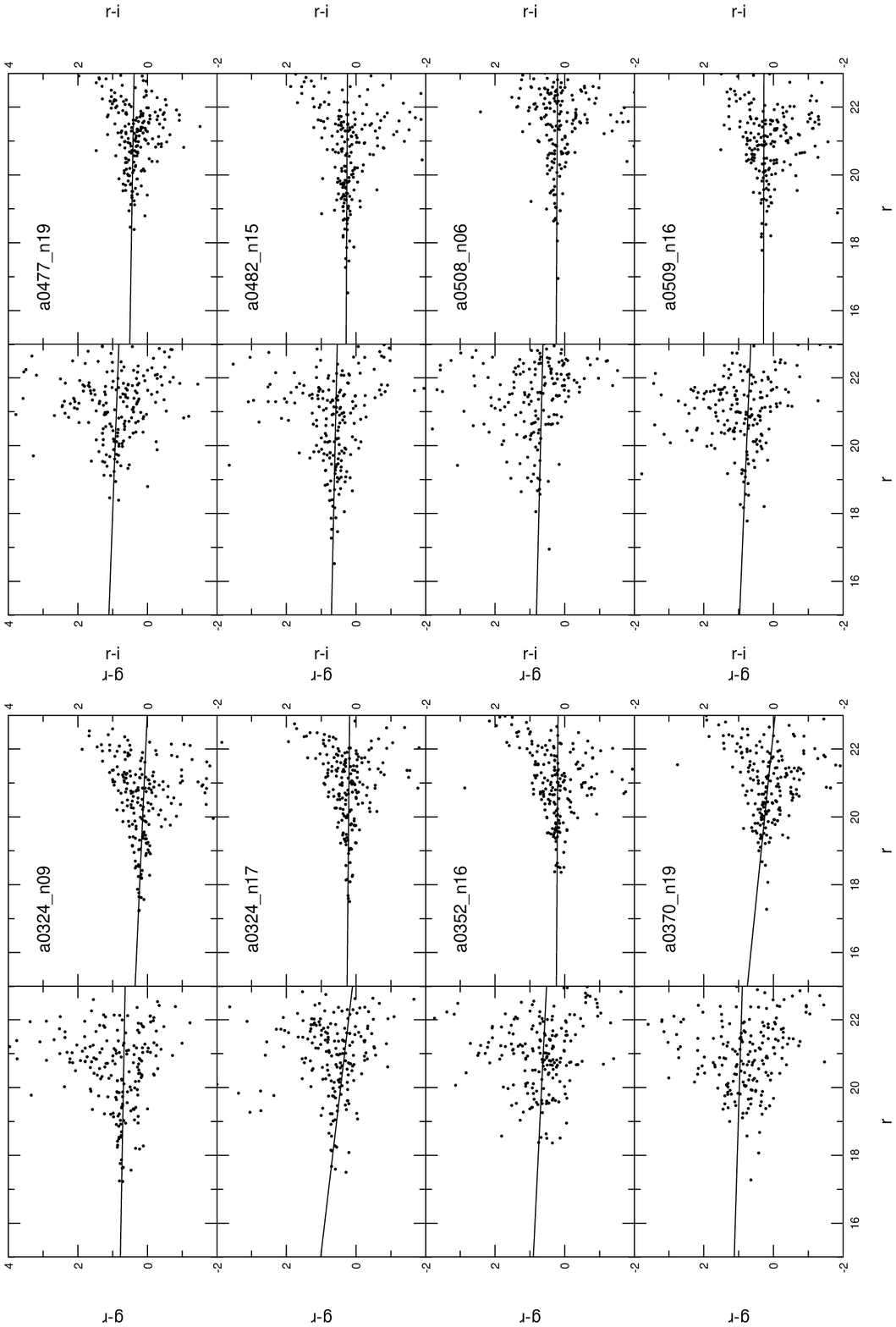}
\caption{Continued.}
\end{figure}

\begin{figure}
\figurenum{3}
\plotone{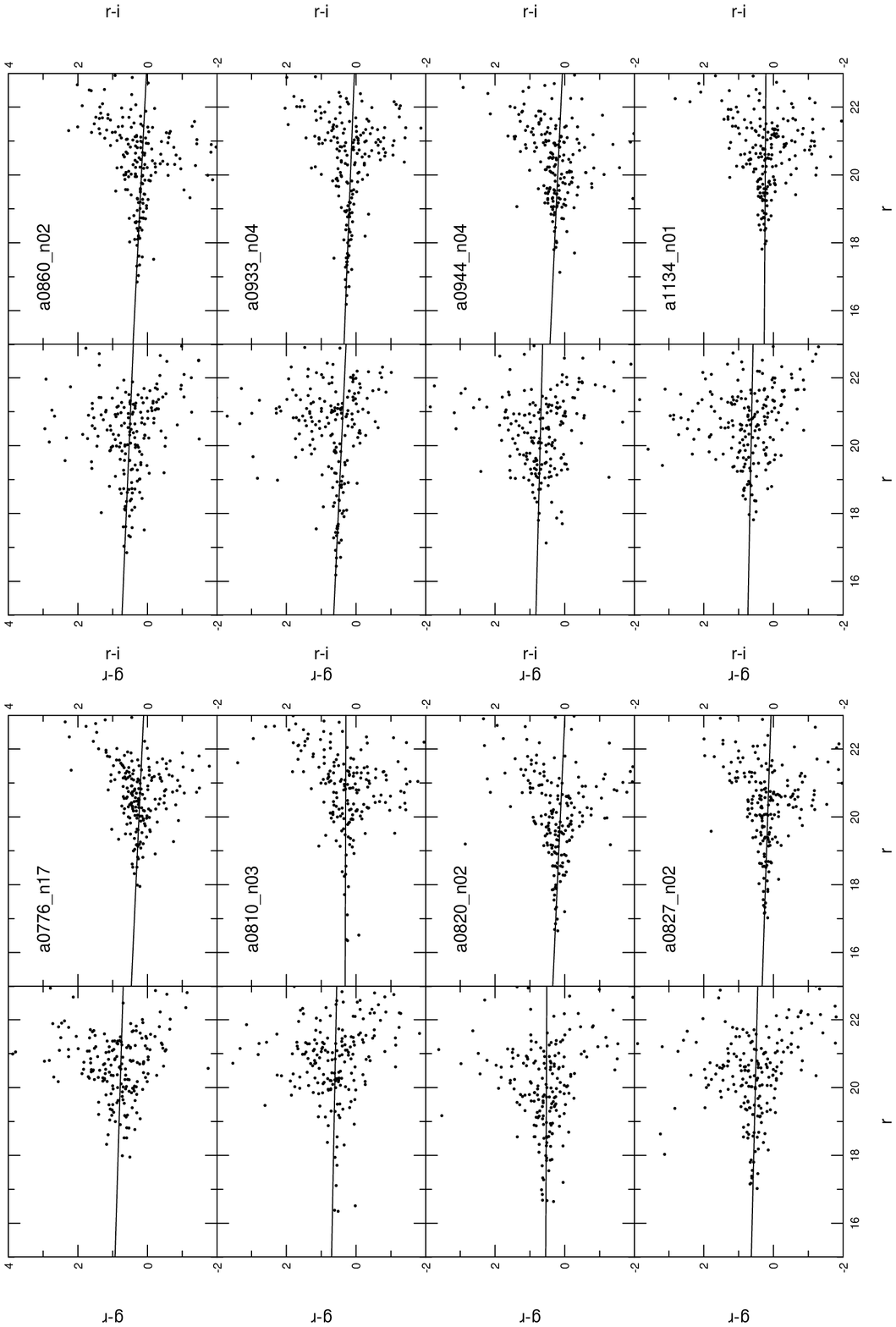}
\caption{Continued.}
\end{figure}

\begin{figure}
\figurenum{3}
\plotone{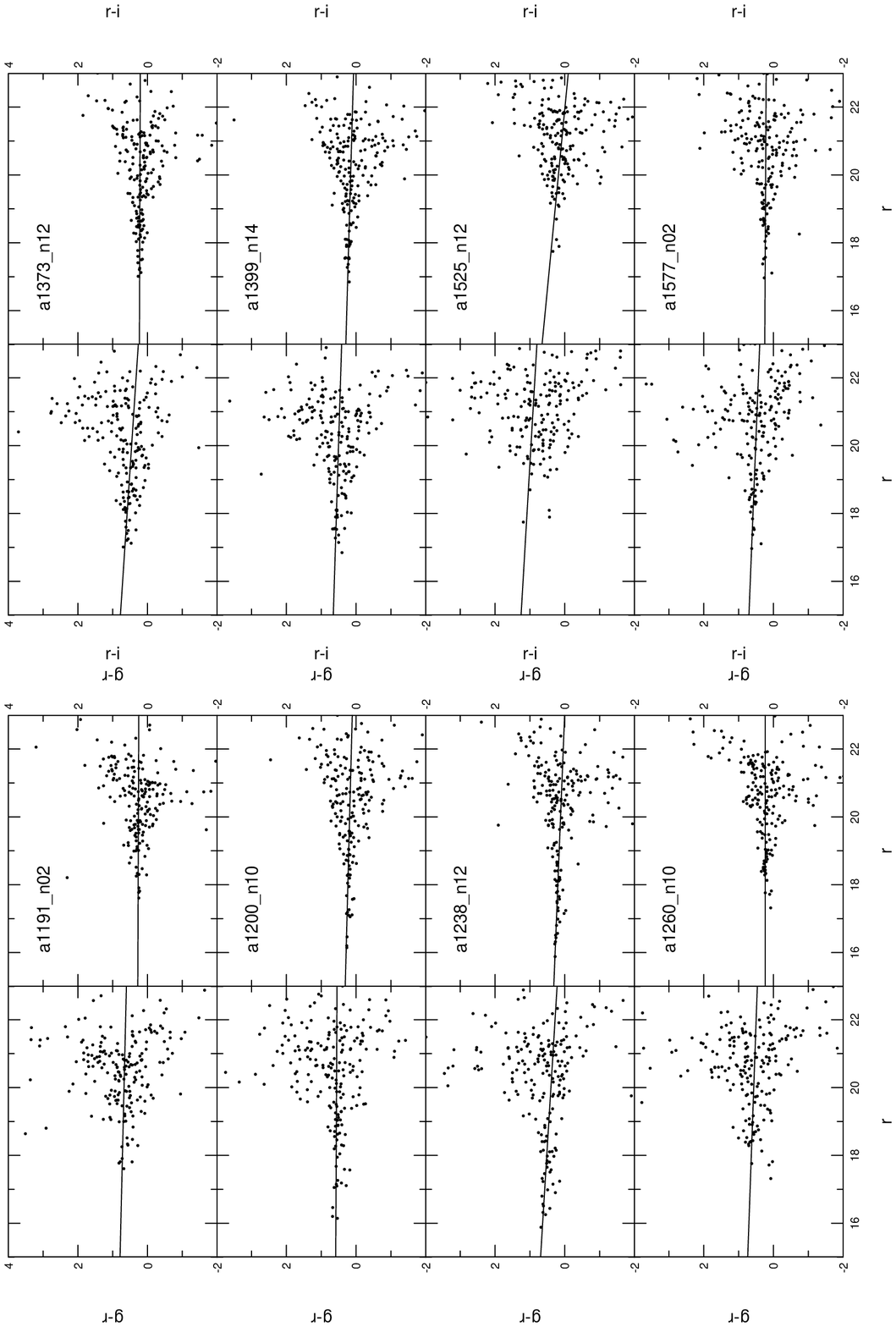}
\caption{Continued.}
\end{figure}

\begin{figure}
\figurenum{3}
\plotone{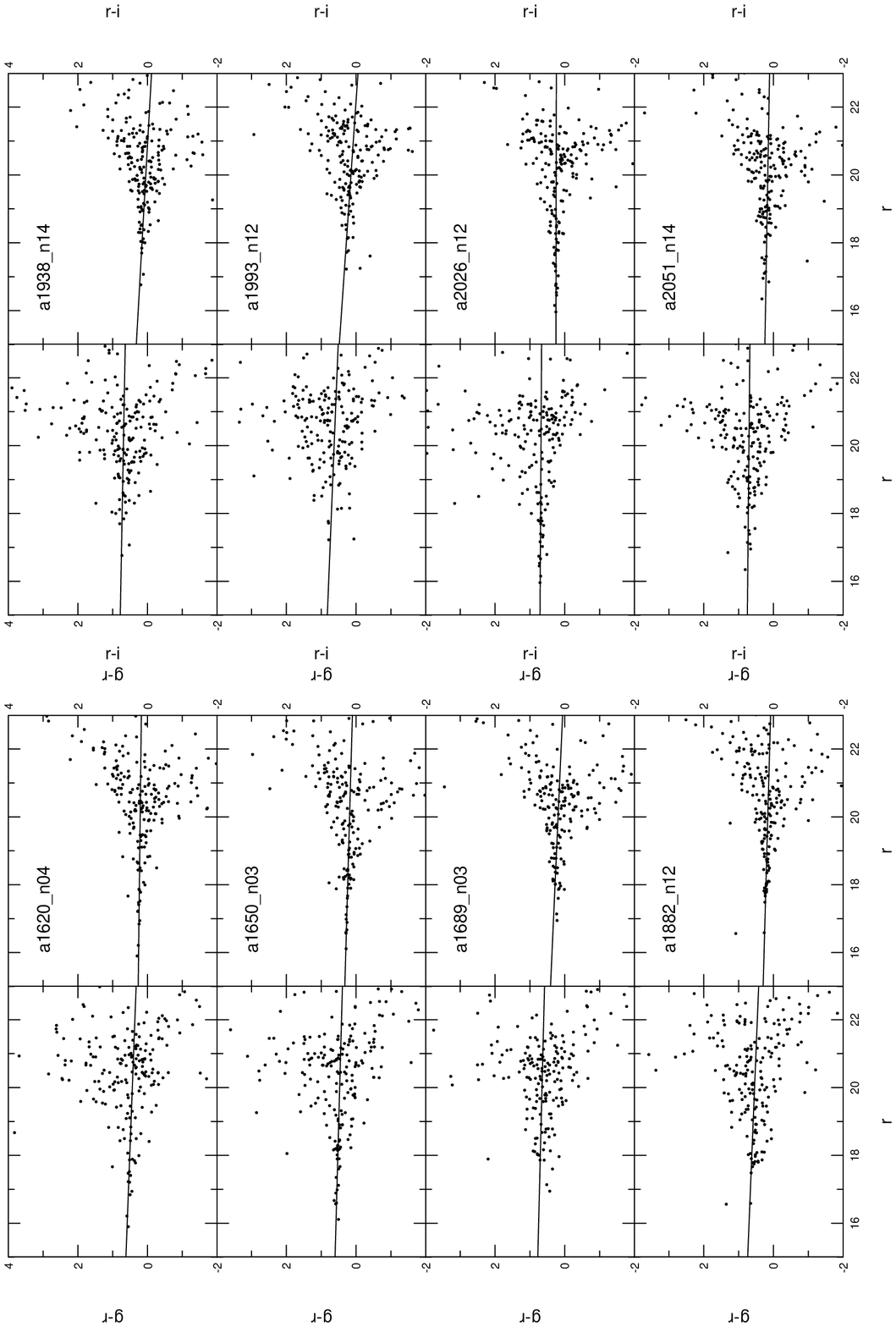}
\caption{Continued.}
\end{figure}

\begin{figure}
\figurenum{3}
\plotone{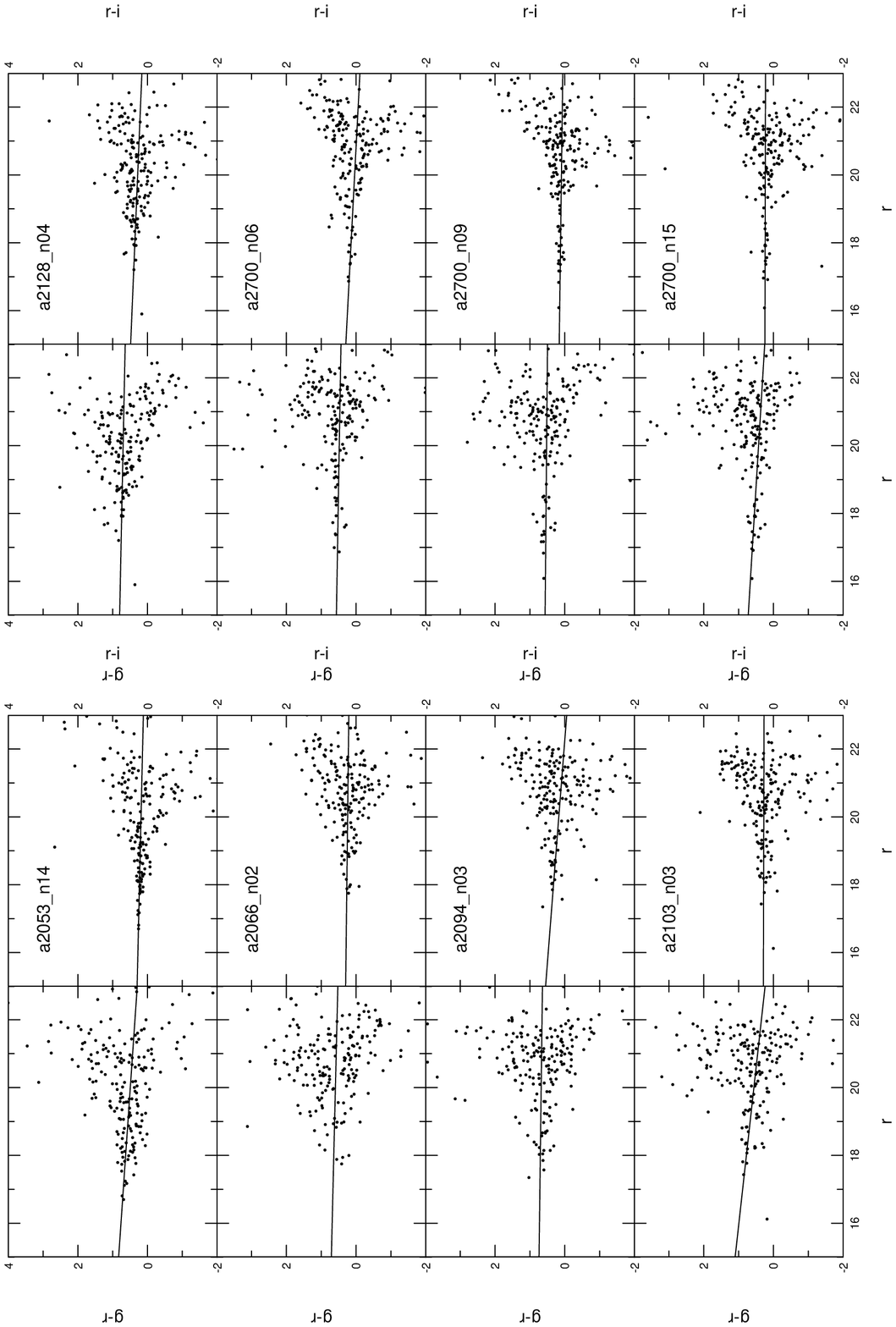}
\caption{Continued.}
\end{figure}

\begin{figure}
\figurenum{3}
\plotone{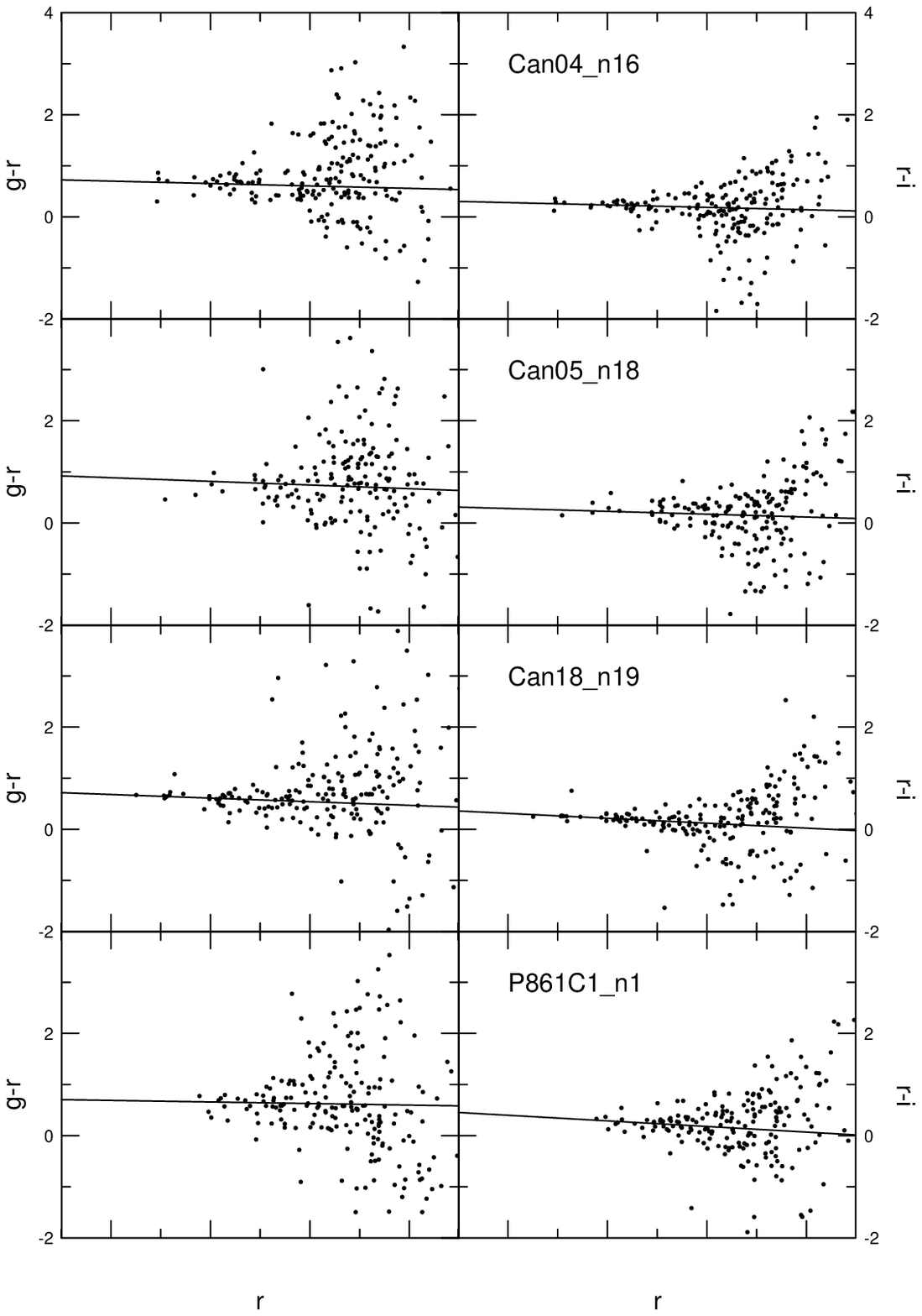}
\caption{Continued.}
\end{figure}

\begin{figure}
\figurenum{4}
\plotone{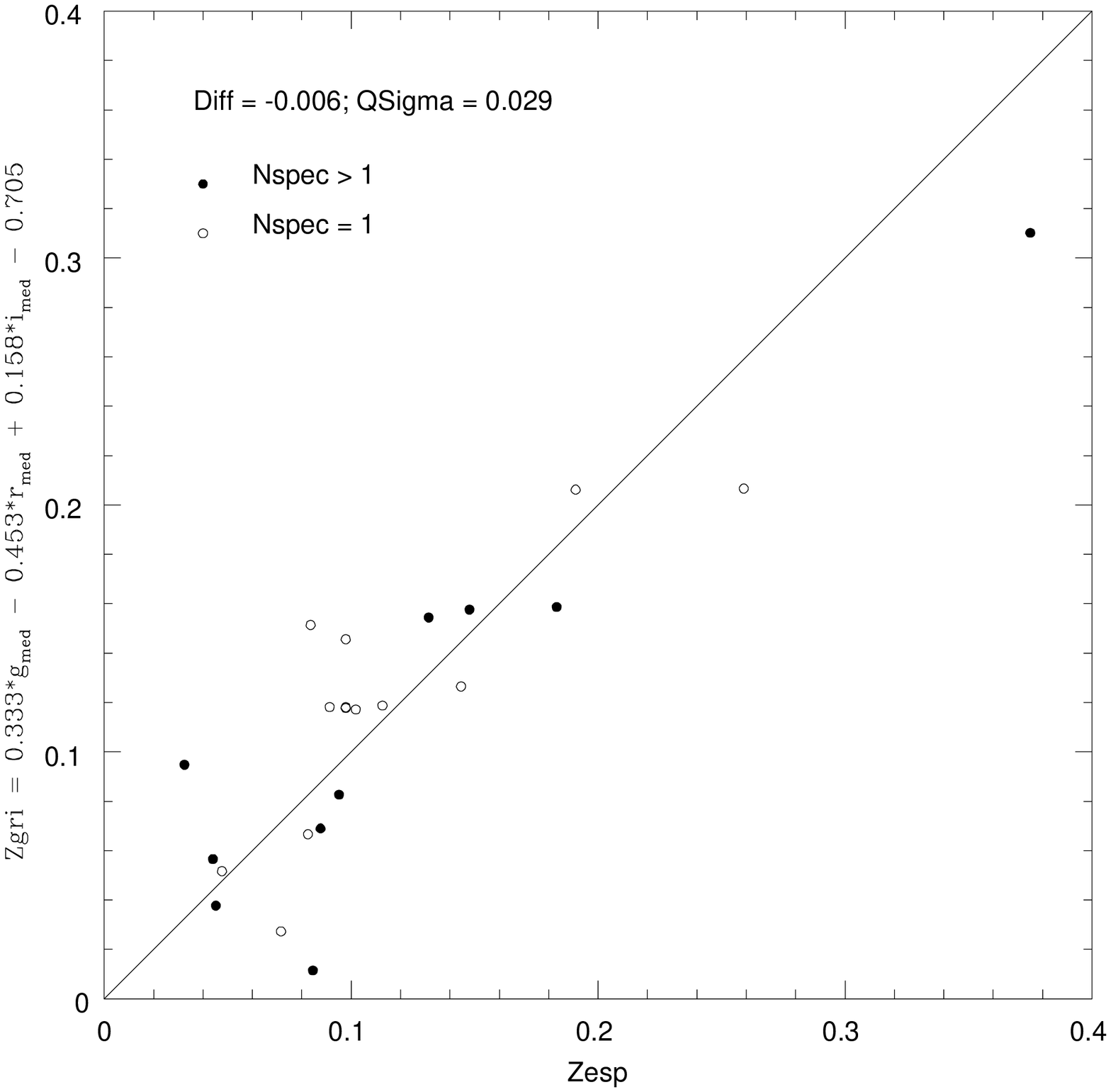}
\caption{Comparison between spectroscopic and photometric
redshift.}
\end{figure}

\begin{figure}
\figurenum{5}
\plotone{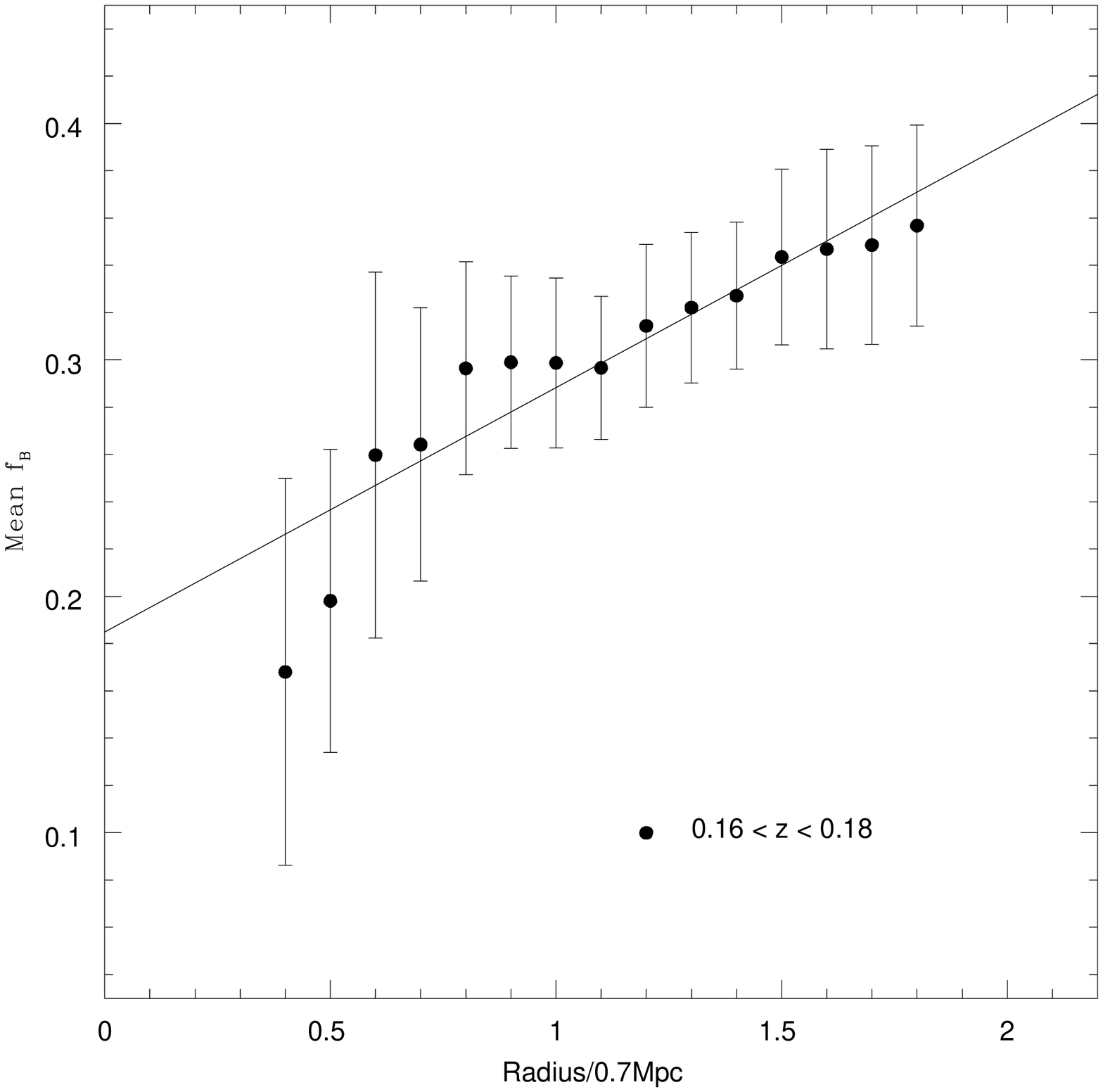}
\caption{Dependence of the fraction of blue galaxies with
the physical region of a cluster.}
\end{figure}

\begin{figure}
\figurenum{6}
\plotone{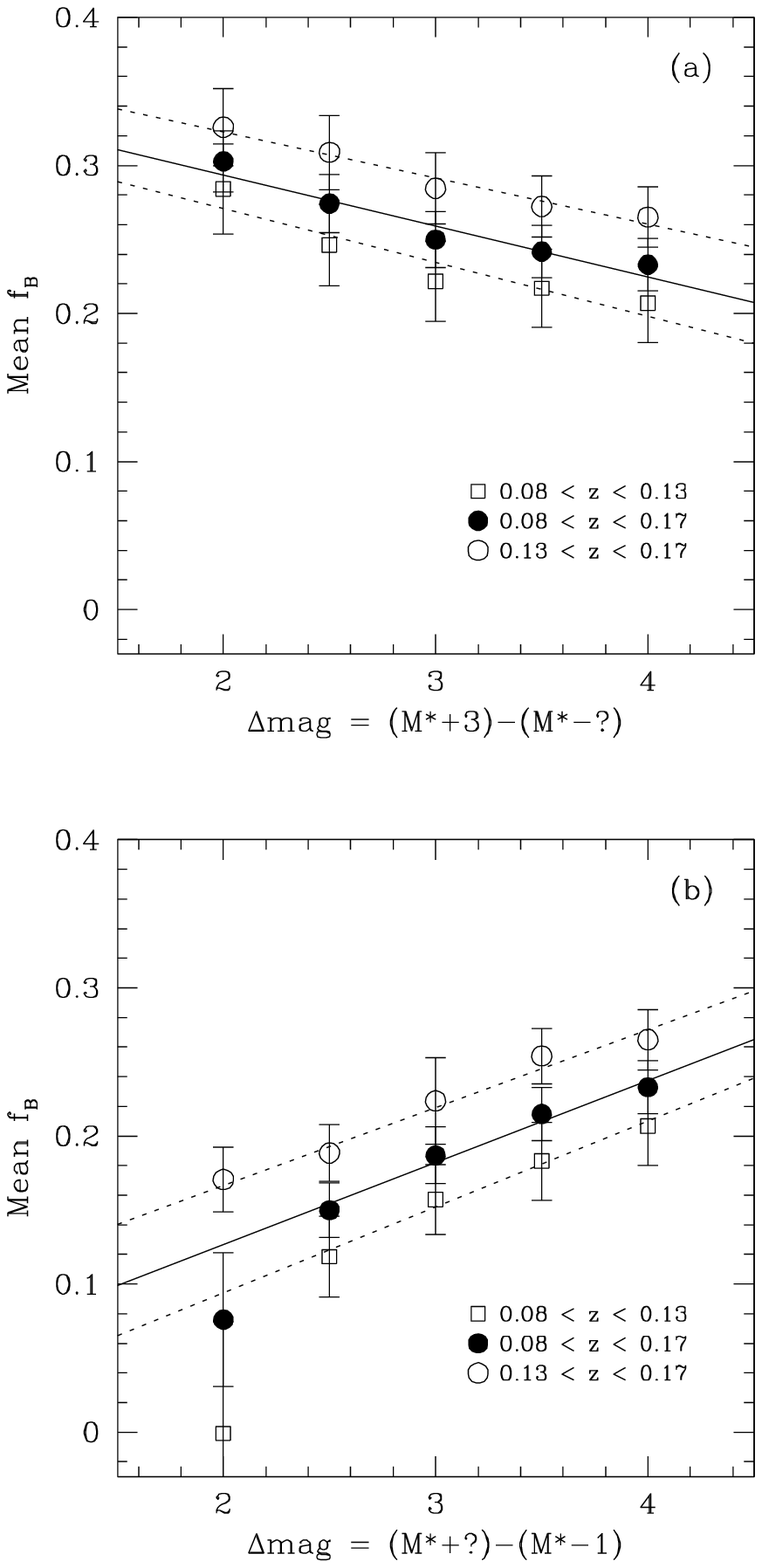}
\caption{Dependence of the fraction of blue galaxies 
with the interval of the luminosity function used to compute it. The 
plot shows the effect of: (a) loosing bright objects and observing only an 
interval $\Delta_{mag}=(M^*+3)-(M^*-?)$; (b) loosing the faintest objects 
and observing only an interval $\Delta_{mag}=(M^*+?)-(M^*-1)$.}
\end{figure}

\begin{figure}
\figurenum{7}
\plotone{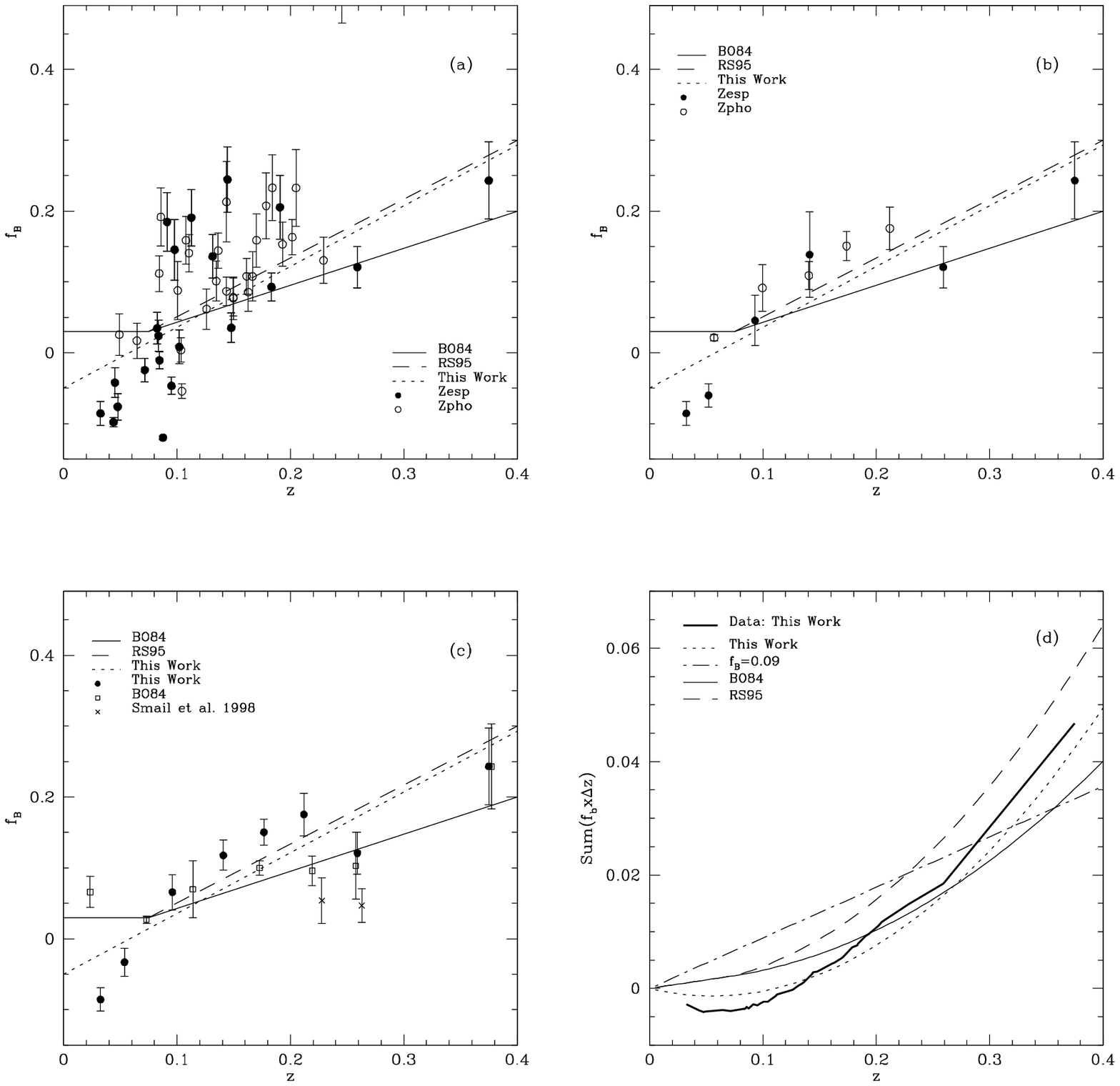}
\caption{(a) Fraction of blue galaxies, with $M^*-1 \le M_{r} \le 
M^*+3$, detected in the central region ($R \le 0.70$Mpc) of the clusters in our sample; 
(b) Comparison between mean $f_B$ values (for intervals of 0.4 in 
redshift) of clusters with spectroscopic and photometric redshifts; (c) Mean 
$f_B$ for BO84, Smail et al. (1998), and for our sample of clusters with 
spectroscopic redshifts; (d) accumulated product of $f_B \times \Delta_z$ for 
the data, BO84 and RS95 relations, and for a constant $f_B$ of 0.10.}
\end{figure}

\begin{figure}
\figurenum{8}
\plotone{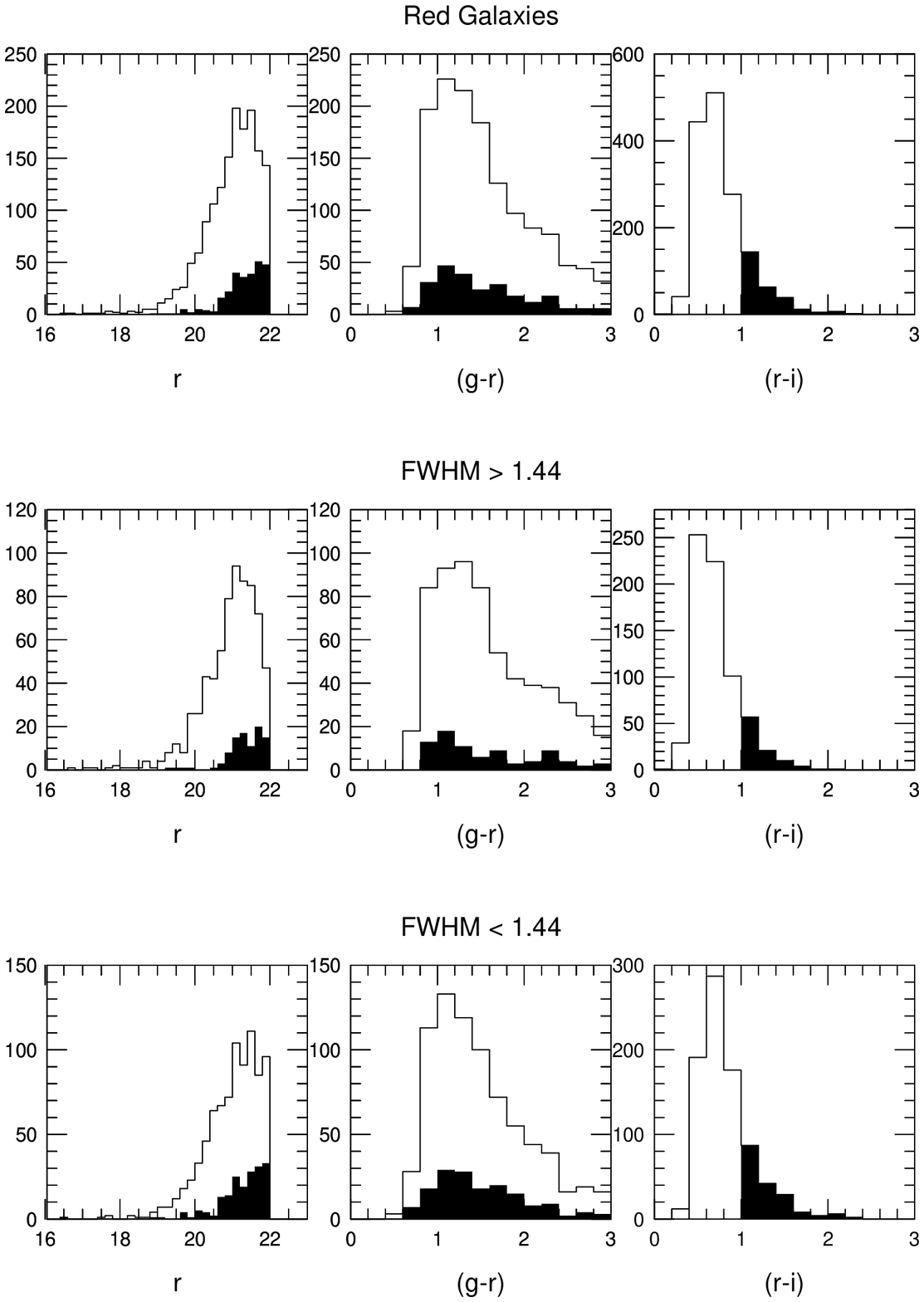}
\caption{Histograms of $m_r$, $g-r$, and $r-i$ for the red galaxies, and for 
the anomalously red galaxied (filled histogram). The upper panels show 
the histograms for all red galaxies detected in our sample. The central 
panels presents the histograms
for the "bad" seeing images ($FWHM > 1.44$), and the "good" seeing images 
($FWHM < 1.44$) are represented in the lower panels. \label{f9}}
\end{figure}

\end{document}